\documentclass[twocolumn]{aastex62}

\received{Aug.30,2020}
\revised{}
\accepted{Nov. 11,2020}
\submitjournal{ApJ}

\shorttitle{X-ray Variability of $\zeta$ Pup}
\shortauthors{Nichols, et al.}
\usepackage{xspace}
\usepackage{amssymb}
\usepackage{times}
\usepackage{verbatim}
\usepackage{color}
\usepackage{graphics}
\usepackage{auto-pst-pdf}
\usepackage{rotating}
\usepackage{graphicx}

\begin{document}

\title{Correlated X-ray and optical variability in the O-type supergiant $\zeta$ Puppis}

\newcommand{\mone}  {^{-1}}
\newcommand{\mtwo}  {^{-2}}
\newcommand{\mthree}{^{-3}}
\newcommand{\mfour} {^{-4}}
\newcommand{\pone}  {^1}
\newcommand{\ptwo}  {^2}
\newcommand{\pthree}{^3}
\newcommand{\pfour} {^4}
\newcommand{\ang}   {{\,\,\AA}\xspace}
\newcommand{\cmmtwo}{\,\mathrm{cm\mtwo}}
\newcommand{\cmptwo}{\,\mathrm{cm\ptwo}}
\newcommand{\cmmthree}{\,\mathrm{cm\mthree}}
\newcommand{\cmpthree}{\,\mathrm{cm\pthree}}
\newcommand{\cts}   {\,\mathrm{cts\,s\mone}}
\newcommand{\cps}   {\,\mathrm{counts\,s\mone}}
\newcommand{\eflux} {\,\mathrm{ergs\,cm\mtwo\,s\mone}}
\newcommand{\kG}    {\,\mathrm{kG}}
\newcommand{\kev}   {\,\mathrm{keV}}
\newcommand{\ev}    {\,\mathrm{eV}}
\newcommand{\kms}   {\,\mathrm{km\,s\mone}}
\newcommand{\pc}    {\,\mathrm{pc}}
\newcommand{\lum}   {\,\mathrm{ergs\,s\mone}}
\newcommand{\mang}  {\,\mathrm{{\,\,\AA}}\xspace}
\newcommand{\mmang}  {\,\mathrm{{m\,\,\AA}}\xspace}
\newcommand{\rmang} {\mathrm{{\,\,\AA}}\xspace}
\newcommand{\rmmang} {\mathrm{{m\,\,\AA}}\xspace}
\newcommand{\mk}    {\,\mathrm{MK}}
\newcommand{\msun}  {\,M_\odot}
\newcommand{\rsun}  {\,R_\odot}
\newcommand{\pflux} {\,\mathrm{photons\,cm\mtwo\,s\mone}}
\newcommand{\apflux} {\,\mathrm{phot\,cm\mtwo\,s\mone}}
\newcommand{\aciss} {{ACIS-S}\xspace}
\newcommand{\acis}  {{ACIS}\xspace}
\newcommand{\ape}   {{\tt acis\_process\_events}\xspace}
\newcommand{\arf}   {{ARF}\xspace}
\newcommand{\cda}   {{CDA}\xspace}
\newcommand{\suz}   {{\it Suzaku}\xspace}
\newcommand{\ciao}  {{CIAO}\xspace}
\newcommand{\cxc}   {{CXC}\xspace}
\newcommand{\cscat} {{CSC}\xspace}
\newcommand{\cxo}   {{\it CXO}\xspace}
\newcommand{\dmcopy}{{\tt dmcopy}\xspace}
\newcommand{\dmex}  {{\tt dmextract}\xspace}
\newcommand{\evto}  {{\tt evt1}\xspace}
\newcommand{\evtt}  {{\tt evt2}\xspace}
\newcommand{\evtz}  {{\tt evt0}\xspace}
\newcommand{\findzo}{{\tt findzo}\xspace}
\newcommand{\heg}   {{HEG}\xspace}
\newcommand{\hetgs} {{HETGS}\xspace}
\newcommand{\hetg}  {{HETG}\xspace}
\newcommand{\hrcs}  {{HRC-S}\xspace}
\newcommand{\hrc}   {{HRC}\xspace}
\newcommand{\isis}  {{ISIS}\xspace}
\newcommand{\ixo}   {{\it IXO}\xspace}
\newcommand{\leg}   {{LEG}\xspace}
\newcommand{\letgs} {{LETGS}\xspace}
\newcommand{\letg}  {{LETG}\xspace}
\newcommand{\meg}   {{MEG}\xspace}
\newcommand{\obs}   {{ObsID}\xspace}
\newcommand{\obss}   {{ObsIDs}\xspace}
\newcommand{\phat}  {{\tt pha2}\xspace}
\newcommand{\rgs}   {{RGS}\xspace}
\newcommand{\rmf}   {{RMF}\xspace}
\newcommand{\slang} {{S-Lang}\xspace}
\newcommand{\tgcat} {{\em TGCat}\xspace}
\newcommand{\tgcm}  {{\tt tg\_create\_mask}\xspace}
\newcommand{\tgdetect}{{\tt tgdetect}\xspace}
\newcommand{\tgextract}{{\tt tgextract}\xspace}
\newcommand{\tgre}  {{\tt tg\_resolve\_events}\xspace}
\newcommand{\hrtennn} {{HR$\,1099$}\xspace}
\newcommand{\hrten}   {\hrtennn}
\newcommand{\hrtenn}  {\hrtennn}
\newcommand{\siggem} {{$\sigma\,$Gem}\xspace}


\newcommand{\Msun}{M$_{\odot}$}
\newcommand{\Rsun}{R$_{\odot}$}
\newcommand{\einstein}{\textit{Einstein}}
\newcommand{\iue}{{\em{IUE}\xspace}}
\newcommand{\nefir}{Ne\,\textsc{ix}}
\newcommand{\neH}{Ne\,\textsc{x}}
\newcommand{\mgfir}{Mg\,\textsc{xi}}
\newcommand{\mgH}{Mg\,\textsc{xii}}
\newcommand{\sifir}{Si\,\textsc{xiii}}
\newcommand{\siH}{Si\,\textsc{xiv}}
\newcommand{\sfir}{S\,\textsc{xv}}
\newcommand{\sH}{S\,\textsc{xvi}}
\newcommand{\ofir}{O\,\textsc{vii}}
\newcommand{\oH}{O\,\textsc{viii}}
\newcommand{\arfir}{Ar\,\textsc{xvii}}
\newcommand{\ergsec}{ergs s$^{-1}$}
\newcommand{\R}{$\mathcal{R}$}
\newcommand{\G}{$\mathcal{G}$}
\newcommand{\RRs}{$R/R_\star$}
\newcommand{\del}{$\delta$}
\newcommand{\vdag}{(v)^\dagger}
\newcommand{\phosec}{ergs s$^{-1}$}
\newcommand{\zp}{$\zeta$ Pup}
\newcommand{\XMM} {{\em XMM-Newton}\xspace}
\newcommand{\Chandra} {{\em Chandra}\xspace}
\newcommand{\BRITE} {{\em BRITE}\xspace}
\newcommand{\IUE}{{\em IUE}\xspace}
\hyphenation{Leeu-wen}
\correspondingauthor{}
\email{jnichols@cfa.harvard.edu}

\author[0000-0003-3298-7455]{Joy S. Nichols}
\affil{Harvard \& Smithsonian Center for Astrophysics, 60 Garden St., Cambridge, MA  02138 USA}

\author[0000-0003-4071-9346]{Ya\"el Naz\'e }
\altaffiliation{FNRS Senior Research Associate}
\affiliation{Groupe d'Astrophysique des Hautes Energies, STAR, Universit\'e de Li\`ege, Quartier Agora (B5c, Institut d'Astrophysique et de Geophysique), All\'ee du 6 Ao\^ut 19c, B-4000 Sart Tilman, Li\`ege, Belgium}

\author[0000-0002-3860-6230]{David P.\ Huenemoerder}
\affiliation{Massachusetts Institute of Technology ,
  77 Massachusetts Ave.,
  Cambridge, MA 02139, USA}

\author[0000-0002-4333-9755]{Anthony~F.~J.~Moffat}
\affil{D\'ept. de physique, Univ. de Montr\'eal, C.P. 6128, Succ. C-V, Montr\'eal, QC H3C 3J7, Canada \& Centre de Recherche en Astrophysique du Qu\`ebec \\}

\author{Nathan Miller}
\affil{Department of Physics and Astronomy, University of Wisconsin-Eau Claire, Eau Claire, WI 54701  USA }

\author{Jennifer Lauer}
\affil{Harvard \& Smithsonian Center for Astrophysics, 60 Garden St., Cambridge, MA  02138 USA}

\author[0000-0002-7204-5502]{Richard Ignace}
\affil{Department of Physics \& Astronomy, East Tennessee State University, Johnson City, TN  37614,  USA }

\author{Ken Gayley}
\affil{Department of Physics and Astronomy, University of Iowa, Iowa City, IA  52242  USA}

\author[0000-0002-8012-0840]{Tahina Ramiaramanantsoa}
\affil{School of Earth and Space Exploration, Arizona State University, 781 E. Terrace Mall, Tempe, AZ, USA 85287-6004}

\author[0000-0003-0708-4414]{Lidia Oskinova}
\affiliation{Institute for physics and astronomy,
  University of Potsdam,
  Karl-Liebknecht-Str. 24/25, 14476,
  Potsdam, Germany
}
\author{Wolf-Rainer Hamann}
\affiliation{Institute for physics and astronomy,
  University of Potsdam,
  Karl-Liebknecht-Str. 24/25, 14476,
  Potsdam, Germany
}

\author[0000-0002-2806-9339]{Noel D.\ Richardson}
\affil{Department of Physics and Astronomy, Embry-Riddle Aeronautical University, 3700 Willow Creek Road, Prescott, AZ 86301, USA}

\author{Wayne L. Waldron}
\affil{Eureka Scientific, Inc.
2452 Delmer Street Suite 100
Oakland, CA 94602 USA}

\author{Matthew Dahmer}
\affil{Northrop Grumman Corporation
2980 Fairview Park Drive
Falls Church, VA 22042 USA}

\thanks{This research makes use of data obtained from the Chandra X-ray Observatory, operated by SAO for and on behalf of  NASA under contract NAS8-03060.}

\thanks{This investigation is based on
data collected by the BRITE-Constellation satellite mission,
designed, built, launched, operated and supported by the
Austrian Research Promotion Agency (FFG), the University
of Vienna, the Technical University of Graz, the Canadian
Space Agency (CSA), the University of Toronto Institute
for Aerospace Studies (UTIAS), the Foundation for
Polish Science \& Technology (FNiTP MNiSW), and National
Science Centre (NCN).}
\NewPageAfterKeywords

\begin{abstract}
Analysis of the recent long exposure \Chandra\ X-ray observation of the early-type O star \zp\ shows clear variability with a period previously reported in optical photometric studies.   These 813 ks of HETG observations taken over a roughly one year time span have two signals of periodic variability:  (1) a high significance period of 1.7820 $\pm$ 0.0008\,d, and (2) a marginal detection of periodic behavior close to either 5\,d  or 6\,d period.  A \textit{BRITE-Constellation} nanosatellite optical photometric monitoring (using near-contemporaneous observations to the \Chandra\ data) confirms a 1.78060 $\pm$ 0.00088\,d period for this star.  The optical period coincides with the new \Chandra\ period within their error ranges, demonstrating a link between these two wavebands and providing a powerful lever for probing the photosphere-wind connection in this star.   The phase lag of the X-ray maximum relative to the optical maximum is $\sim\phi$=0.45, but consideration of secondary maxima in both datasets indicates possibly two ``hot" spots on the star with an X-ray phase lag of $\phi$=0.1 each.  The details of this periodic variation of the X-rays are probed by displaying a phased and trailed X-ray spectrum and by constructing phased light curves for wavelength bands within the HETG spectral coverage (ranging down to bands encompassing groups of emission lines).  We propose that the 1.78\,d period is the stellar rotation period and explore how  stellar bright spots and associated co-rotating interacting regions (CIRs) could explain the modulation of this star's optical and X-ray output and their phase difference.

\end{abstract}


\section{Introduction}

Massive stars have significant impacts on the abundances, evolution, and energy budgets of the galaxies they inhabit, both through the powerful stellar winds they produce during their lifetimes, and through the supernova explosions which are often their ultimate fate.   A complete understanding of the physical mechanisms involved in  massive star evolution and their associated winds is still elusive, but their ubiquitous variations can provide an important probe to the underlying mechanisms. Variability of massive stars has been observed on various timescales and with diverse amplitudes, from giant ejections of matter in unstable evolutionary stages (i.e., Luminous Blue Variables) to small-scale stochastic wind variations.

In this context, \zp\  is a key target of study, as it is the closest, at a distance of 332$\pm$11\, pc \citep{howarth19}, and one of the brightest O-type supergiants, having a spectral type O4Inf  \citep{sota14}.   The intense scrutiny of this star has led to the detection of several types of variability in multiple wavelengths. First, photospheric optical absorption-line profile variations with an 8.5\,h period were reported by \citet{baade86} and \citet{reid96}, and interpreted in terms of non-radial pulsations, but the variations appear transient (e.g. Baade 1991). In the ultraviolet (UV), cyclic variability attributed to Discrete Absorption Components (DACs) was detected \citep{kaper99}.   Between 1989 and 1995 that variability increased in apparent period from  $\sim$15\,h to $\sim$19\,h. This change in period in the UV was detected by the International Ultraviolet Explorer \IUE\ UV \citep{howarth95}, H$\alpha$ \citep{reid96}, and ROSAT X-rays \citep{berghoefer96}.
 A still longer period of about 5\,d was detected in H$\alpha$ and \IUE\ UV lines \citep{moffat81,howarth95} and was proposed to be associated with rotation; but again, repeatability proved to be elusive.

More recently, \citet{howarth14} reported photometric changes with a 1.78\,d period using the Solar Mass Ejection Imager (SMEI) instrument on the Coriolis satellite. The same period was
subsequently confirmed in the 5.5 month BRIght-star Target Explorer (\BRITE)  campaign in 2014--2015 by \citet{ramia18} and is still seen in subsequent data up to the present (Ramiaramanantsoa et al., in prep.). Both sets of data from optical space photometry are completely dominated by continuum light from the stellar photosphere. While being of stable period, the shape of the phased 1.78d
light curve is not sinusoidal and was seen to change on timescales of weeks or months during the 2014--2015 \BRITE\ campaigns.

Nearly simultaneous cyclical variations of He\,\textsc{ii} $\lambda$4686\,\AA\ and other emission lines with the same 1.78\,d period were also found from observations in parallel with the \BRITE\ campaign \citep{ramia18}. These show a phase lag which increases to $\sim \phi$=0.1 for the optical emission lines formed furthest from the star.
Although \citet{howarth14} interpret the 1.78\,d periodicity variations in terms of non-radial pulsations, \citet{ramia18} link the
changes to slowly changing bright spots on the surface potentially driven from below by subsurface convection.  The photometric \BRITE\ campaign also demonstrated the presence of stochastic photometric changes of similar amplitude to the periodic variation on the 1.7806\,d period.  These stochastic variations are only coherent on a multi-hour timescale, and are suspected to arise in the photosphere from the same subsurface activity related to a zone of partial ionization of iron-group elements \citep{cantiello09}.

Analysis of a 10-year $\sim$ 1Ms \XMM\ dataset failed to detect significant short-term changes on the order of hours beyond Poisson statistics, thus indicating a highly fragmented wind \citep{naze13}.   The \XMM\ observations also pointed towards the presence of larger, slow modulations of the X-ray flux, with peak-to-valley amplitudes of $\sim$15\%\ over the duration of the individual exposures (typically $\sim$16\,h, \citealt{naze13}). These changes appeared strongest in the  \XMM\ medium-energy (0.6--1.2\,keV; equivalent in energy to the \Chandra\ soft) band \citep{naze18}.

X-ray emission in massive stars is thought to be generated from the natural instabilities of the line-driven stellar wind \citep{lucy80,cass83,feldmeier97}. Hot stars have shown a variety of  periodic temporal behaviors in their X-ray emission, with some displaying connections between different wavelength regimes.
A recent study \citep{Massa19} compared \XMM, STIS, and \IUE\ data of $\xi$ Per, finding a consistent 2.086\,d period in the datasets and a small time lag that is dependent on the ionization state of each line.
CIR relationships were proposed for the O supergiant $\lambda$ Cep \citep{rauw15} and the dwarf $\zeta$ Oph \citep{oskinova01}.

In this paper we report a detailed analysis of the 2018--2019 \Chandra\ observing campaign using the Advanced CCD Imaging Spectrometer (ACIS) with the High Energy Transmission Grating (HETG) to characterize the X-ray variability of \zp.  By including in our analysis the near-simultaneous optical \BRITE\ data for this star, we can explore links between these two wavelength bands.  Different physical regions are causing the emission in these two wavelength bands, so any connections found between them will have ramifications for understanding the connection between the star's photosphere and its outflowing wind.     The observational data used in these analyses
are described in Sect.\,\ref{sec:obs}.  We began by constructing and analyzing full-band X-ray light curves for all of the HETG data.
Potential periods were identified from these light curves.  We analyzed these X-ray data and the nearly-simultaneous optical data in concert with one another to probe for connections between them (Sect.\,\ref{sec:lightcurve}).
Light curve analyses of specific wavelength regions for emission lines  are presented in Sect.\,\ref{sec:spectral}. Then, the entire dataset was partitioned into short time intervals of calibrated data, and the variability of lines and continuum were examined using moments (Sect.\,\ref{sec:time_sliced_spectra}).
Sect.\,\ref{sec:discussion} is a discussion and interpretation of the results, and the conclusions are
presented in Sect.\,\ref{sec:conclusions}.

\section{Observations and Data Analysis} \label{sec:obs}

The \Chandra\ observations of $\zeta$ Pup discussed here, with a total exposure time of 813 kiloseconds (ks), used the HETG with  the ACIS-S instrument \citep{canizares05}.  The gratings provide separate Medium Energy Grating (MEG) and High Energy Grating (HEG) spectra simultaneously, with resolutions of 0.023\,\AA\ and 0.012\,\AA\, respectively.
Table \ref{tab:obs} lists each  observation (\obs) acquired for this program, and the exposure time.

Spacecraft thermal considerations required most of the observations to use 4--5 CCD chips instead of the full array of 6 chips, truncating the spectral range in the longer wavelengths.  There is also one early 2000 \Chandra\ HETG observation of \zp, \obs\ 640, that we did not include in this analysis because it was acquired too many cycles ago (18--19
years) to be useful in the context of phasing with a short periodicity due to the accuracy of the ephemeris.

%
\begin{deluxetable}{rrr}
\tabletypesize{\scriptsize}
 \tablecaption{List of X-Ray observations divided into summer 2018, winter 2019, and summer 2019 observation groups}
 \tablehead{
   \colhead{ObsID}&
   \colhead{Observation Start Time\tablenotemark{a}}&
   \colhead{Exposure (ks)}
 }
 \startdata
21113& 2018-07-01T20:18:49& 17.7  \\
21112& 2018-07-02T22:57:54& 29.7  \\
20156& 2018-07-03T16:06:38& 15.5  \\
21114& 2018-07-05T17:00:36& 19.7  \\
21111& 2018-07-06T05:00:09& 26.9  \\
21115& 2018-07-07T03:17:11& 18.1  \\
21116& 2018-07-08T02:20:58& 43.4 \\
20158& 2018-07-30T22:36:40& 18.4  \\
21661& 2018-08-03T11:42:46& 96.9  \\
20157& 2018-08-08T23:32:35& 76.4  \\
21659& 2018-08-22T02:13:29& 86.3  \\
21673& 2018-08-24T18:52:10& 15.0  \\
\hline
20154& 2019-01-25T03:21:34& 47.0 \\
22049& 2019-02-01T00:55:26& 27.7 \\
\hline
20155& 2019-07-15T00:04:38& 19.7  \\
22278& 2019-07-16T16:20:37& 30.5  \\
22279& 2019-07-17T14:52:40& 26.0  \\
22280& 2019-07-20T06:45:30& 25.5  \\
22281& 2019-07-21T21:13:28& 41.7  \\
22076& 2019-08-01T00:47:34& 75.1 \\
21898& 2019-08-17T03:16:06& 55.7  \\
 \enddata
 \tablenotetext{a}{Terrestrial Time}
\end{deluxetable}\label{tab:obs}

\begin{figure*}[htb]
\centering
\includegraphics[width=\textwidth,viewport=2 285 612 512,clip=true]{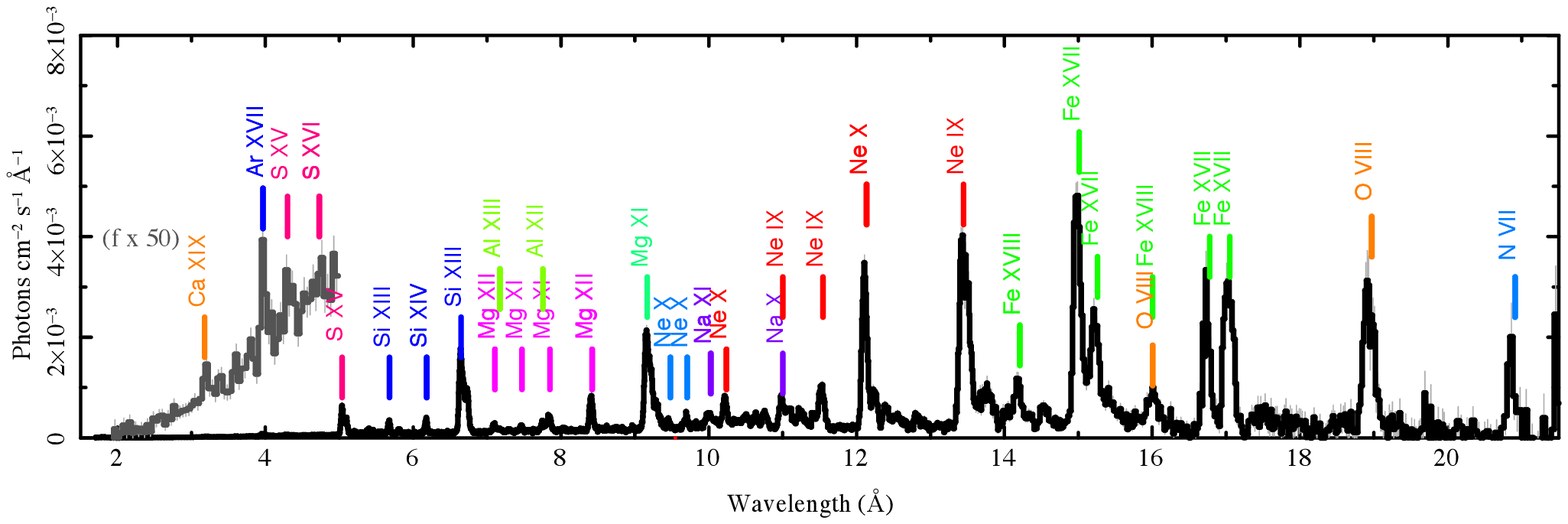}
 \label{fig:spectrum}
 \caption{X-ray spectrum over the entire waveband of \Chandra\ HETG ACIS-S including all 813\,ks of exposure time taken in 2018 and 2019.  Fluxes shortward of 5\,\AA\ have been multiplied by a factor of 50 to improve legibility.  The highest-energy lines and the high-energy bremsstrahlung continuum were analyzed in \citet{huene20}.    The spectrum principally consists of lines of Fe\,\textsc{xvii}, adjacent ionization states of iron, and hydrogen- and helium-like ions of Ne, Mg, S, and Si and O.}
\end{figure*}

Each observation in Table \ref{tab:obs} was processed using the TGCat software \citep{Huenemoerder11}, starting with the Level 1 (bias-corrected, unfiltered) event data.  TGCat software uses CIAO \citep{Frusc06} tools to process the data, yielding filtered Level 2 event data, extracted spectra, and appropriate calibration files.
Fig.\,\ref{fig:spectrum} shows the cumulated spectrum of all observations, along with major emission line identifications.

It is well-known that the \Chandra\ ACIS optical blocking filter has a
buildup of molecular contamination, still increasing with time, which
degrades the X-ray transmission especially at lower
energies\footnote{For details of ACIS contamination, see the
  \it{\Chandra\ Proposers' Observatory Guide}, \S 6.5.1, \url{
    https://cxc.harvard.edu/proposer/POG/html/chap6.html\#tth\_sEc6.5.1}}.
To remove this known effect from count-rate light curves, we evaluated
the expected count rates in each band by adopting the mean \zp\ flux as a model spectrum and folding this through the
responses at the epoch of each observation.  The mean flux provides an
appropriate weighting function vs energy within each band.  We fit a
linear function to these model rates and used the slope to remove this
instrumental trend from the light curves (correct\_ct\_rate = original\_ct\_rate - slope * (HJD-HJD\_1st\_obs))
before conducting any timing
analysis.  See Table \,\ref{tab:bands} for specific values used in each bandpass.

\section{Broad-band light curves and Periodicity}\label{sec:lightcurve}

\subsection{Broad-band light curves}

Broad-band light curves were constructed for each ObsID, then combined into a single light curve for the entire \Chandra\ campaign consisting of count rates per 4\,ks bin.  Counts from HEG and MEG were combined where their spectra overlap in wavelength (3.1--10.3\,\AA).   Then the MEG counts in the range 10.3--20.7\,\AA\  were  added to this total, providing a total wavelength coverage of 3.1--20.7\,\AA.    The break at 10.3\,\AA\ was chosen because some of our \obss\ did not have HEG coverage beyond this position.  This break in wavelength coverage also provides compatible ranges to \XMM\ hard and \XMM\ medium bands (note the definitions of ``hard" and ``medium" differ between \XMM\ and \Chandra; see Table \ref{tab:bands}).   In this paper we will refer to the \Chandra\ definitions of hard, medium, and soft, and an additional band that we call ``hybrid-hard" that refers to all of the \Chandra\ medium band and part of the \Chandra\ hard band, in order to directly compare to the \XMM\ hard band.  All times in the light curve were corrected to the barycentric times.  The bins are contiguous and exclusive, which means there is always a short time bin at the end of each observation.  We binned to 4\,ks intervals uniformly from the beginning of each observation.   Short end bins ($\le$50\% exposure per bin) were excluded from the timing analysis due to their larger variance which could degrade results.  A total of six short time bins were excluded, containing less than 1.5\% of the total exposure.

As in \citet{naze18}, we first estimated the amount of variability in the
light curves using the indices VI and $F_\mathrm{var}$ \citep [see Appendix A] {ede02}. VI is a variability index, expressed as
$(max-min)/(max+min)$, with max and min values taken  without any exclusion and thus making them prone to noise fluctuations; it thus compares the amplitude of the count rate
variation (i.e. half the peak-to-peak variation amplitude) relative to the mean.
The more useful fractional variability amplitude $F_\mathrm{var}$ provides an idea of the amplitude of {\it intrinsic} changes, relative to the mean, as it is defined as $$F_\mathrm{var} = \sqrt{S^2-\sigma^{2}_\mathrm{err}}/X_\mathrm{m},$$ with the mean $X_\mathrm{m}=\sum{X_{\mathrm{i}}}/N$, $i$=bin index, the dispersion $S^2=\sum(X_{\mathrm{i}}-X_{\mathrm{m}})^{2}/(N-1)$ and the mean Poisson error $\sigma_{\mathrm{err}}^{2}=\sum{{\sigma_{\mathrm{err,i}}^{2}}/N}$.  This  eliminates the variance that is due to Poisson noise \citep{ede02}.
A comparison of the variability in \Chandra\ data and \XMM\  using these indices is presented in Table \ref{f_var}.   Note that both \Chandra\ and \XMM\ values agree with each other and, as found in \XMM\ data (Naze et al. 2018), the 10.3-20.7\,\AA\ band appears more variable than the 3.1-10.\,\AA\ band in \Chandra\ data (see Table \ref{tab:bands})..

\begin{table}
\scriptsize
\centering
\caption{Comparison of \Chandra\ and \XMM\ $F_\mathrm{var}$}
\label{tab:f_var}
\setlength{\tabcolsep}{3.3pt}
\begin{tabular}{rrr}
\hline\hline
Mission & Wavelength & $F_\mathrm{var}$    \\
\hline

\Chandra\ &10.3--20.7\AA\ & 0.039$\pm$0.009 \\
\XMM\ & 10.3--20.7\AA & 0.040$\pm$0.002 \\
\Chandra\ & 3.1--10.3\AA\ & 0.030$\pm$0.005 \\
\XMM\ & 3.1--10.3\AA\ & 0.027$\pm$0.002 \\
\hline
\end{tabular}
\label{f_var}
\end{table}

\subsection{Period Search}\label{sec:period}

\begin{figure*}[htb]
  \centering
  \includegraphics[width=\textwidth]{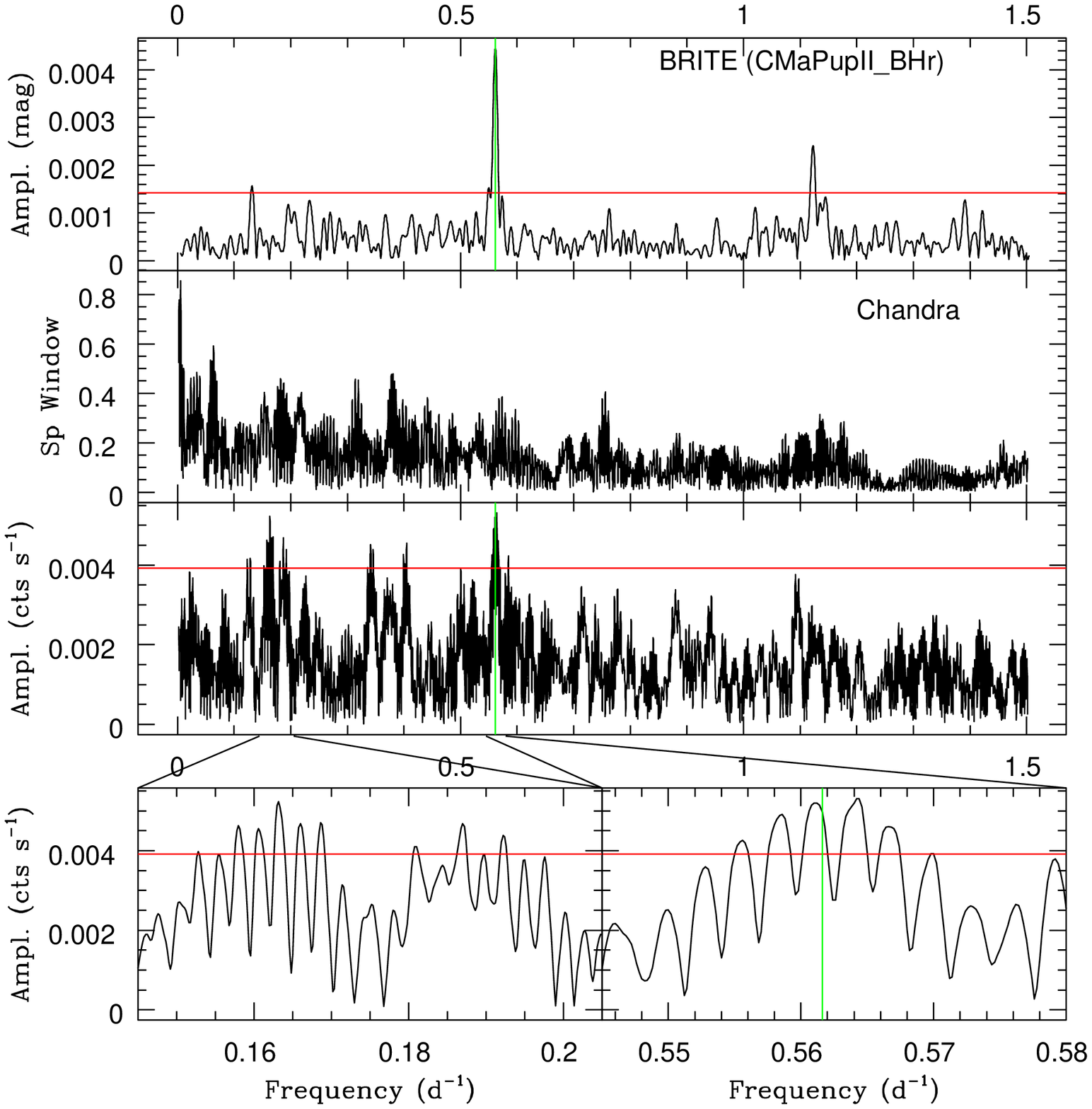}
 \caption{Periodogram of the broad-band corrected \Chandra\ light curve of \zp, along with its spectral window and zooms on two important regions discussed in text. The red horizontal line represents the 1\% significance level while the green vertical lines correspond to the period detected in optical data \citep{ramia18}. For comparison, the top panel provides the periodogram for the 2018-2019  \BRITE\ observing campaign, nearly contemporaneous with the \Chandra\ run.}
\label{fig:yael}
\end{figure*}

 The corrected \Chandra\ light curves were analyzed using a modified Fourier algorithm adapted to datasets with uneven sampling \citep{heck85,gosset01,zech09}\footnote{Three other types of period-search methods were employed on the data: (1) analyses of variances (e.g. AOV,
\citep{sch89}), (2) conditional entropy \citep{graham13,cin99a,cin99b}, and (3)  Lomb-Scargle periodogram.  These methods gave similar results to the modifed Fourier method used.}.   Fig.\,\ref{fig:yael}  shows the obtained periodogram with a red horizontal line representing the 1\% significance level. This level was estimated by two methods, which agree with one another: (1) by performing 2,000 Monte-Carlo simulations drawing at random the individual count rates from a Gaussian distribution with the same mean for all simulations and a standard deviation equal to the dispersion of the observed count rates, and (2)  by shuffling the data. The maximum amplitudes reached by the periodograms of the simulated data were recorded and the significance level was fixed at the amplitude above which only 1\% of the periodogram maxima lie.  This 1\% significance level thus provides the amplitude which will be detected in a dataset similar as ours by chance; some may consider it as an upper limit because the simulations include the effects of real variations (two periodicities along with all their many harmonics and aliases), which will enhance the dispersion of the observed count rates beyond that of the Poisson noise level.

 Because the sampling of the \Chandra\ observation is very irregular, many aliases are present in the periodogram (see the spectral window in Fig.\,\ref{fig:yael}). In particular, the dataset is composed of short snapshots spread over a year, so any single peak will appear as a broad compound of narrow subpeaks (see zoomed images at the bottom of Fig.\,\ref{fig:yael}).
 In the periodogram,  two groups have the highest significant frequencies in units of cycles per day (c/d): near 0.17\,c/d$^{-1}$ and near 0.56\,c/d$^{-1}$. Since the second one has the largest amplitude, we first focus on it.   A sum of 10 gaussians were fit to the periodogram over the region of the strongest peaks (lower right panel in Fig.\,\ref{fig:yael}).   Of the 10 peaks fit, the peak with centroid 0.561147\,c/d, indicating a period of 1.7820\,d, was chosen as the closest to the \BRITE\ period and therefore the most likely true period.

  We estimated  the error in the period to be of the order of 1/10T=1/(10$\times$411)\,d = 0.00024 c/d, with T being the total time interval of all the \Chandra\ observations in days (as in Ramiaramanantsoa, et al., in prep.).  The method yielded a period error of 0.0008\,d.  We checked this error value using the Gaussian fits described above, where the uncertainty in the centroid value, the sigma value, and the HWHM  are consistent with 10\% of the peak width and with our adopted error value.

The highest peak at $P =1.7727 \pm0.0008$\,d within this group of subpeaks is not within the errors of the X-ray period of 1.7820\,d and is probably due to the irregular sampling; occasionally in such cases, a subpeak or alias (rather than the highest peak) corresponds to the actual signal.  The presence of nearly the same periodicity in both optical and X-ray datasets clearly points towards a common origin for both variability phenomena.

\begin{table}
\scriptsize
\centering
\caption{Ephemerides determined  from X-ray and optical data}
\label{tab:ephem}
\setlength{\tabcolsep}{3.3pt}
\begin{tabular}{lll}
\hline\hline
Source of data & Period & T$_0$ \\
\hline
\Chandra & $P =1.7820 \pm 0.0008 $\,d  & \dotfill \\
BRITE & $P=1.7806 \pm 0.00088$\,d & HJD=2,458,425.800 \\
\hline
\end{tabular}
\end{table}

\begin{figure}[htb]
  \centering
  \includegraphics[width=.5\textwidth,viewport=0 0 230 365,clip=true]{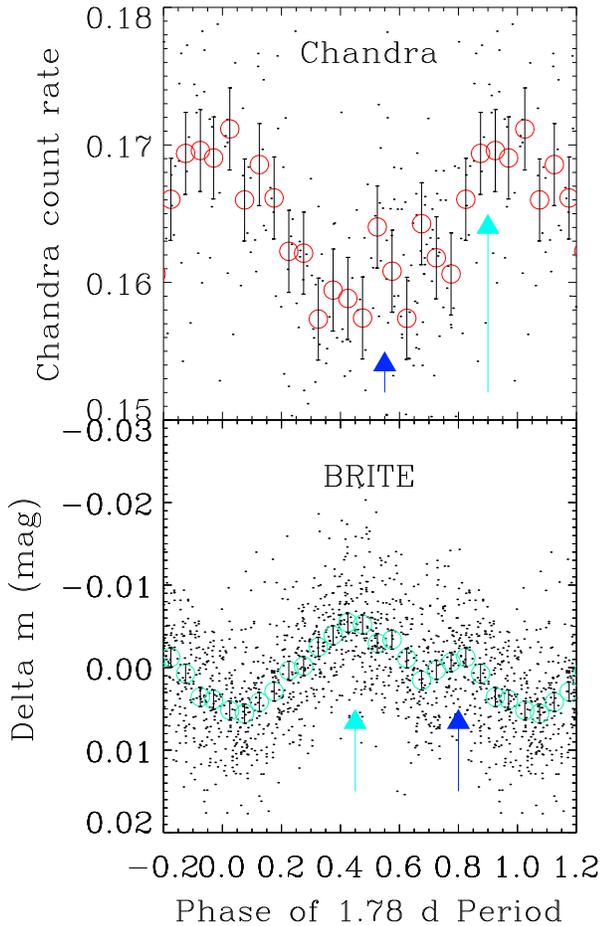}

  \caption{Folded optical and X-ray light curves.  Top: Broad-band X-ray light curve, with data binned at 4\,ks as small black dots and more coarsely binned light curve data at 0.05 phase as larger red circles.  Error bars are standard error on the mean for each binned point.   Each point represents the mean of the time bins that fall in the phase bin.  The number of time bins included in a data point vary from 5 to 21 due to the uneven phase coverage.  One full cycle is shown, with an additional 0.2 in phase replicated at each end to show continuity with phase. Cyan arrows indicate the maximum of each light curve.  Blue arrows indicate the secondary maximum of each light curve.  Bottom: \BRITE\ data with the mean subtracted. Binned light curve data at 0.05 phase are shown as green circles. Error bars are standard error on the mean for each binned point.}
  \label{fig:joy}

  \end{figure}

Given that the optical \BRITE\ value is consistent from several years of data \citep[see notably][]{ramia18}, we adopt the \BRITE\ ephemeris (period $P=1.7806$\,d and $T_0$ is HJD=2,458,425.800) for any calculation of light curve phases related to this period. Using this ephemeris places the optical maximum at phase 0.45,  The X-ray light curves are always shown with that ephemeris, in order to compare the light curves directly.  Figure \ref{fig:joy} shows both X-ray and optical light curves folded with this ephemeris. For the binned X-ray light curve, we calculated averages of data points in 20 phase bins over the 1.7806\,d period. Each phase bin is $\phi$=0.05.  The \BRITE\ light curve has been binned in the same manner.   The peak-to-valley amplitude of this binned light curve is about 6\%, which is remarkably large compared to the optical variations (peak-to-valley amplitude of about 1\%).

The light curves in Fig.\ref{fig:joy} are notably for several features.  The minimum of the X-ray light curve corresponds to the maximum of the optical light curve, approximately.  The \BRITE\ light curve has a clear maximum at $\phi$=0.45 as well as a secondary maximum at about $\phi$=0.8.  The \Chandra\ light curve has a maximum at $\sim\phi$=0.9 and a secondary maximum at about $\sim\phi=0.55.$  The phase of the primary X-ray maximum was determined by cross-correlation between the \BRITE\ light curve and the \Chandra\  light curve, yielding an offset of 0.45 in phase.    The secondary X-ray maximum is based on the "point" at 0.55 in phase, but this point represents the mean of 18 time bins from 12 observations that fell in this phase bin after folding on the 1.78 day period.  Each point in Fig.\,\ref{fig:joy} was calculated in the same way, but the number of time bins that fell in each phase bin varies from 5 to 21 due to the uneven coverage of the observations.  The data point for the secondary X-ray maximum in Fig. \ref{fig:joy}  is calculated to be somewhat more than 2$\sigma$ from the mean of the residuals after fitting the light curve with a 2-degree polynomial. Although this is a marginal detection, it nevertheless should be explored and is discussed in Sect. \ref{lag}.  A nearby point at about $\phi$=0.65 has approximately equal count rate as the  peak at $\phi=0.55$, but is less than 1.5$\sigma$ from the mean of the residuals.

\begin{figure}[!htb]
\centering
\includegraphics[width=0.4\textwidth,viewport=0 0 539 699,clip=true]{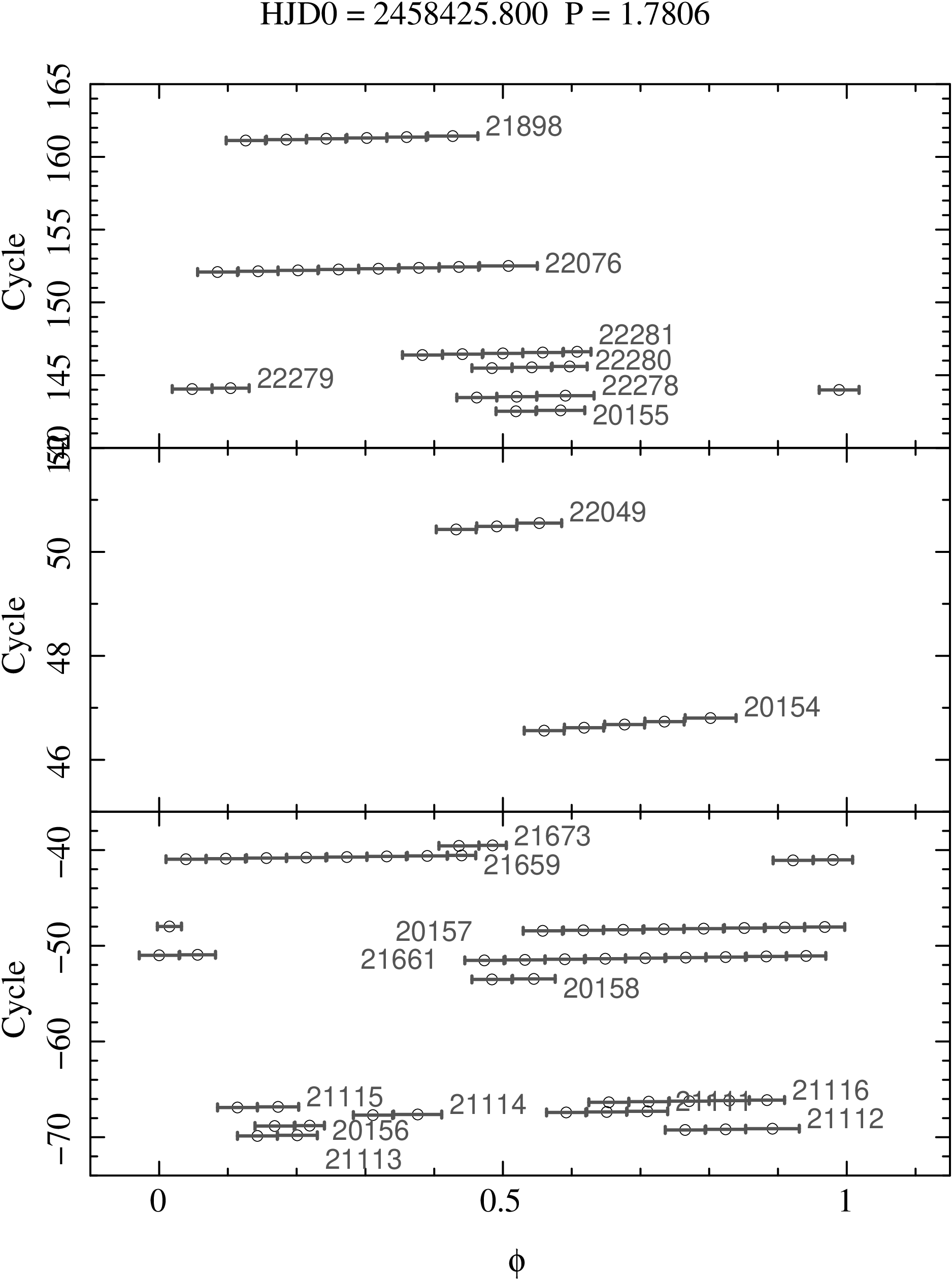}
\caption{Diagram showing phase coverage of \Chandra\ \zp\ observations, assuming phases based on the ephemeris of the prominent 1.7806\,d period (see text), as a function of cycle number of the same period with zero-point in the middle of the whole data-set. }
 \label{fig:cycle}
\end{figure}

\begin{figure}[htb]
  \centering
  \includegraphics[width=.5\textwidth,viewport=0 0 386 410,clip=true]{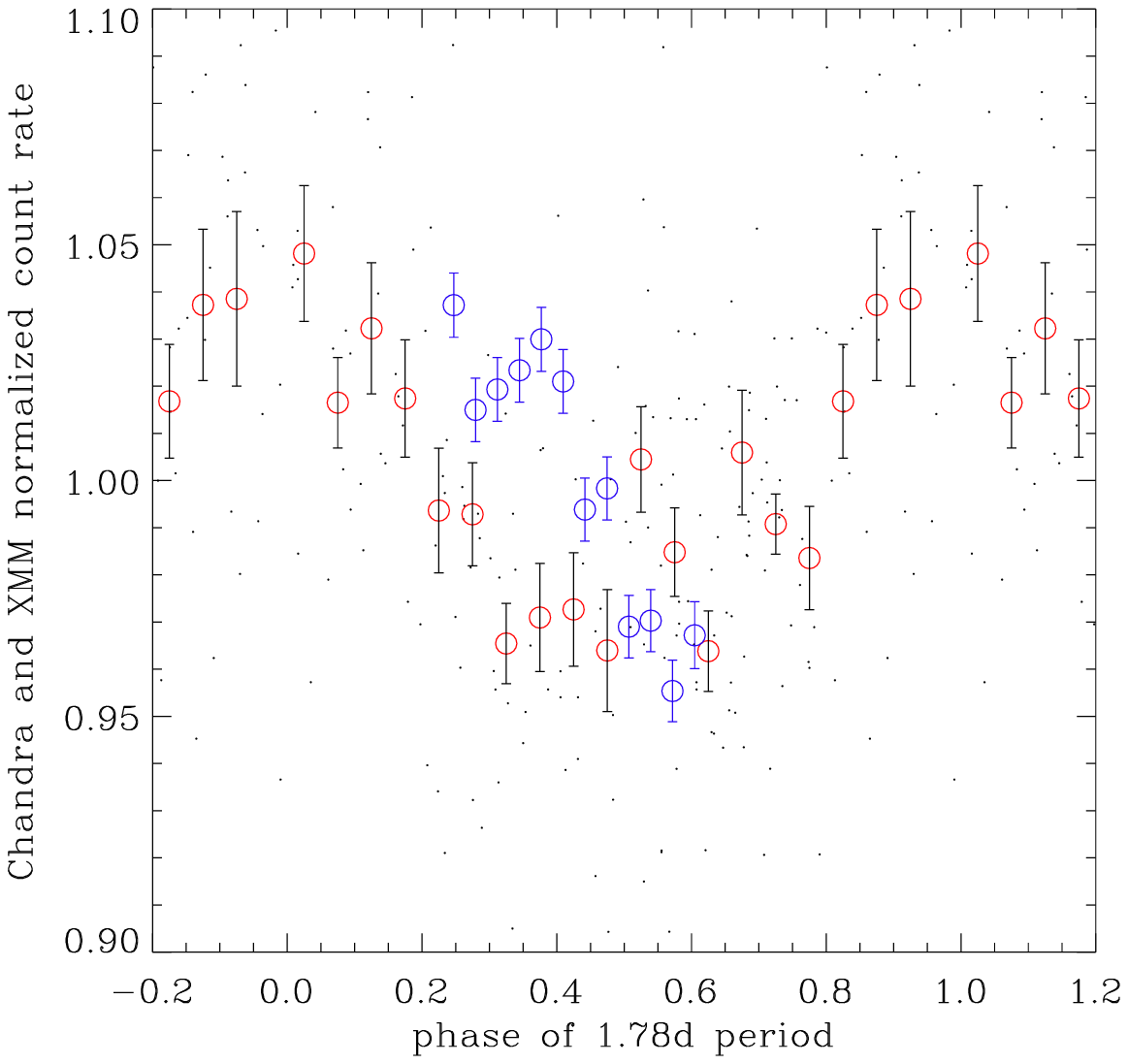}

  \caption{Folded and normalized \Chandra\ and \XMM\ data on 1.7806\,d period.  \XMM\ data (blue circles) and errors are from April 2019.  \Chandra\ data (red circles) as in Fig.\ref{fig:joy}.  The \Chandra\ data were normalized to the \Chandra\ data mean and the \XMM\ data were normalized to the \XMM\ data mean.}
  \label{fig:joy_xmm}
  \end{figure}

 Once we applied the same period and ephemeris to the \BRITE\ data and \Chandra\ data reported here,
the peaks of these two light curves did not coincide. The maximum X-ray emission lags by $\Delta(\phi)\sim0.45$ behind the maximum optical emission, a value determined by cross-correlation of the two light curves.
 Such a large time lag  would place constraints on the connections between the optical and X-ray emission regions, as will be discussed in more detail in Sect.\,\ref{sec:discussion}.  Alternately, if we take both the primary and secondary maxima in the optical light curve and relate them to the secondary and primary X-ray maxima, we have two optical events with X-ray lags time of $\sim \phi$=0.1 each.

When taken as a whole, this \Chandra\ data set for \zp\ provides coverage of all phases of a 1.78\, period.  However, not all phases are covered equally well in all observation segments.  This can lead to some limitations.  For example, an important question is whether the period detections are stable over time. At first glance, since the \Chandra\ data have been mostly obtained in two observing windows, summer 2018 and summer 2019, they could allow for such a check. Unfortunately, as shown by the observation coverage (Fig.\,\ref{fig:cycle}), the maximum of the 1.7806\,d modulation was not sampled during the second observing window. However, it is important to note that the binned light curves of the two observing windows, when they do overlap in phase, appear compatible, favoring the hypothesis of the periodogram stability.

Fig.\,\ref{fig:joy_xmm} compares the \XMM\ calibration observation taken in April 2019 (blue circles) to the \Chandra\ data.  The \XMM\ minimum  agrees approximately with the \Chandra\ minimum. The light curve evolution appears different, with a steeper declining slope for the \XMM\ data. However, one needs to keep in mind that while the string of \XMM\ data points in the figure represent a single exposure, the \Chandra\ big dots represent data points at these phases from a combination of the whole campaign.   One would not expect exact agreement between \XMM\ and \Chandra\ because \XMM\ data are sensitive to  softer X-rays than \Chandra\ data, and because of the difference in sensitivity between the \XMM\ and \Chandra\ instruments.  Looking at the small dots of Fig.\,\ref{fig:joy_xmm}, which represent the individual \Chandra\ 4\,ks bins, we see that the \XMM\ binned light curve appears well within the scattered \Chandra\ points.

\begin{figure}[!htb]
\begin{tabular}{cc}
\includegraphics[width=3.in,viewport=0 0 571 548,clip=true]{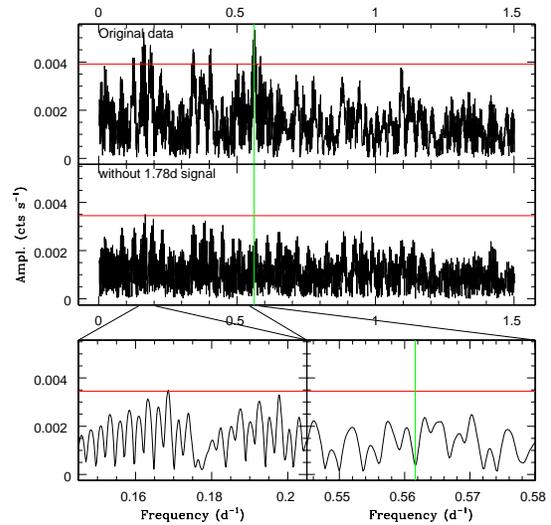}
 \end{tabular}
\caption{Comparison of the initial periodogram and the periodogram once the light curve has been cleaned by the 1.7806\,d signal. Green vertical lines as in Fig.\,\ref{fig:yael}}
\label{fig:prewhit_fourier}
\end{figure}

To further inquire about the presence of additional coherent signals or of stochastic variability beyond the 1.7806\,d signal,  we once again calculated averages of data points, but this time in 10 phase bins rather than 20; the choice of 10 phase bins per cycle is a compromise between having enough signal in each bin while still allowing us to examine the shape of the phased light curve in detail.
Linear interpolations of this binned \Chandra\ light curve were used to remove from the \Chandra\ data the variations associated with the 1.7806\,d period.  Because the light curve does not appear to be a perfect sine wave  (see Fig.\,\ref{fig:prewhit_fourier}), there must be harmonic content in addition to the fundamental.  While an improved data cleaning for this period would take into account the contributions of these additional harmonic components, our periodogram (Fig.\,\ref{fig:yael}) does not provide enough information on the harmonic content to allow for this.  Hence the choice of the subtraction of the mean light curve.

We then performed a period search on the resulting cleaned light curve (Fig.\,\ref{fig:prewhit_fourier}). The overall peak amplitudes appear largely reduced in this periodogram, indicating that the 1.7806\,d signal dominates the X-ray variability. However, one peak is still significant, at a frequency of 0.16860$\pm$0.0002\,d$^{-1}$, corresponding to a period of 5.9312$\pm$0.009\,d. The next largest peak, only slightly lower signficance, lies at $0.1978\pm 0003$\,d$^{-1}$, corresponding to a period of 5.056$\pm$0.008\,d. It is actually difficult to choose which one of those two peaks corresponds to the ``real'' signal.  In fact, different processing choices may lead to one or the other peak leading in the cleaned periodograms. This indicates that, most probably, a period of about 5\,d or about 6\,d is present.

\begin{figure*}[!htb]
\begin{tabular}{cc}
\includegraphics[width=3.in]{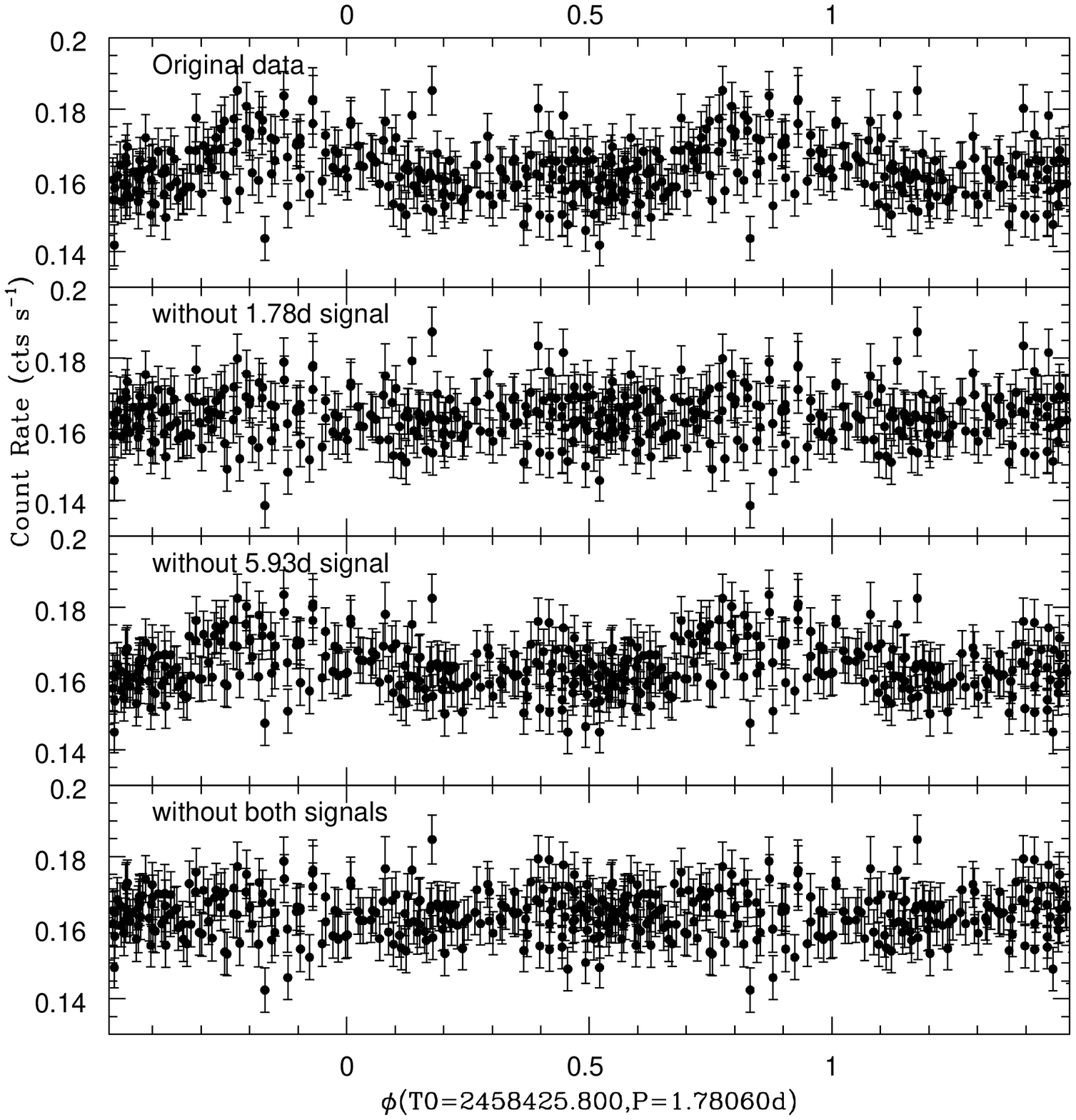}  &
\includegraphics[width=3.in]{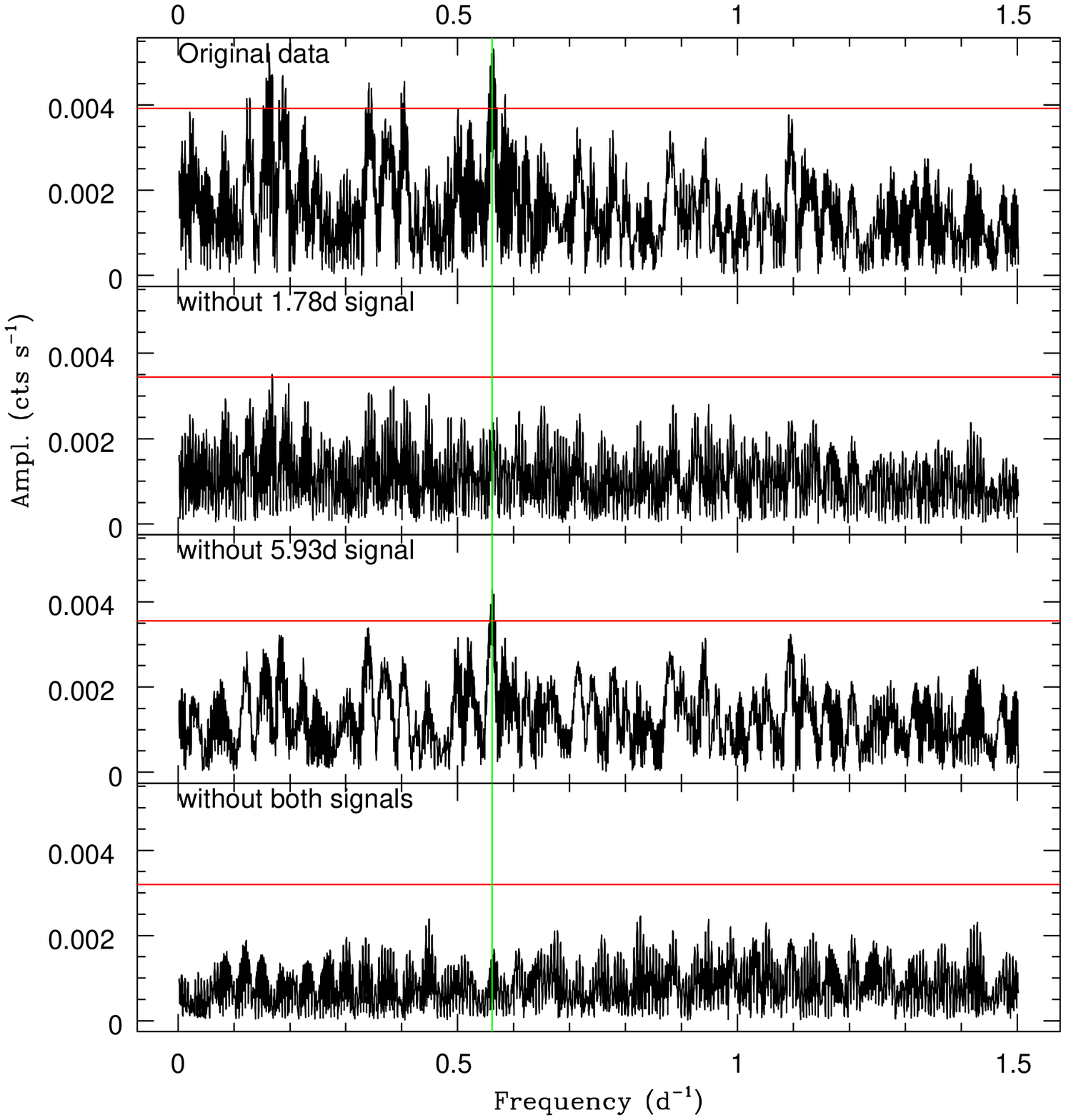} \\
 \end{tabular}
\caption{Left: \Chandra\ light curves folded with the optical ephemeris at different stages of cleaning: initial curve on top panel, curve after cleaning by the 1.780\,d signal on second panel, curve after cleaning by a 5.9312\,d signal on third panel, curve after cleaning by both signals on bottom panel.
Two cycles are shown; data have been replicated, in order to show continuity with phase.  Right: Periodograms at different levels of cleaning.  Top panel provides the periodogram of the initial light curve, the second panel shows the periodogram after cleaning for the 1.7806\,d signal, the third panel shows the periodogram after cleaning for a 5.93\,d signal, and the bottom panel presents the resulting periodogram after cleaning by both signals. As before, the red horizontal line represents the 1\% significance level.}
\label{fig:prewhit_fourier2}
\end{figure*}

Focusing on the formally significant period only, the corrected light curve, cleaned for the 1.7806\,d signal, was folded with a 5.9312\,d period and averages in 10 phase bins were then calculated. The original light curve was then cleaned by this average 5.9312\,d signal in the same manner as done before for the 1.7806\,d case. The periodogram of this newly cleaned light curve is shown in Fig.\,\ref{fig:prewhit_fourier2}. It is immediately obvious that the 1.7806\,d signal remains significant. It is thus important to note that both signals are distinct, i.e. even though they may be close to a harmonic ratio of 3, they appear separate as cleaning by one does not remove the other. As a last trial, we cleaned the original light curve by both averaged curves and performed a period search on the result (Fig.\,\ref{fig:prewhit_fourier2}). This time, the periodogram amplitudes are very much reduced and no significant signal is detected.

As a final exercise, we calculated the fractional variability amplitudes $F_\mathrm{var}$ of the light curves at different stages of cleaning (Sect.\ref{sec:lightcurve}).  $F_\mathrm{var}$ was 0.031$\pm$0.004 for the original light curve; it decreased to 0.020$\pm$0.005 after cleaning for the 1.7806\,d signal, or 0.024$\pm$0.004 after cleaning for the 5.9312\,d signal, and finally 0.011$\pm$0.007 after cleaning by both signals. The latter value implies that zero (i.e. no variability beyond Poisson noise) is only at 1.2$\sigma$.  Hence the observed variability of the \Chandra\ light curve can be explained {\it a posteriori} by two periodic signals (1.7820\,d and 5 or 6\,d), with the presence of additional variations, stochastic or coherent, to be confirmed (or excluded) with higher-quality data in the future.

\begin{figure}[htb]
  \centering
  \includegraphics[width=.5\textwidth,viewport=0 0 386 410,clip=true]{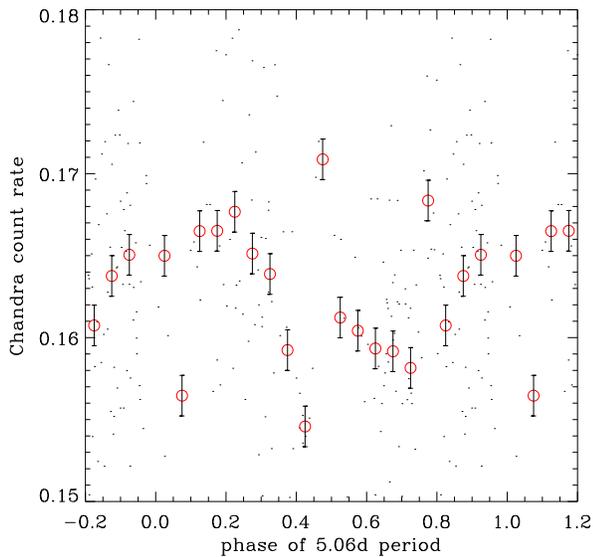}

  \caption{Folded X-ray light curves on a period of 5.06\,d.   Broad-band X-ray light curve, with data binned at 4\,ks as small black dots and more coarsely binned light curve data at 0.05 phase as larger red circles.  Error bars are standard error on the mean for each binned point.  One full cycle is shown, with additional 0.2 in phase replicated at each end to show continuity with phase.}
  \label{fig:joy_506}
\end{figure}

We also plotted in Fig.\,\ref{fig:joy_506} the \Chandra\ light curve folded on a 5.06\,d period which was previously proposed by \citet{moffat81}.  This light curve is possibly consistent with the presence an $\sim$ 5\,d period, although there are several significant outliers.  We discuss this potential period in Sect.\,\ref{sec:discussion}.

\begin{table*}
\scriptsize
\centering
\caption{Limits of the energy bands used for the broad-band lightcurves.  "m" indicates that a band include only MEG data.}
\label{tab:bands}
\setlength{\tabcolsep}{3.3pt}
\begin{tabular}{lccccc}
\hline\hline
Name & limits (\AA) & $VI$ & $F_\mathrm{var}$ & slope (cts\,s$^{-1}$\,d$^{-1}$) & Remarks\\
\hline
full &  3.1--20.7 & 0.13$\pm$0.03 & 0.031$\pm$0.004 & -2.660e-05 & 0.6--4.0\,keV,m \\
\Chandra\ soft band (=\XMM\ medium)   & 10.3--20.7 & 0.25$\pm$0.05 & 0.039$\pm$0.009 & -1.829e-05 & 0.6--1.2\,keV,m \\

\Chandra\ hybrid hard band (=\XMM\ hard)   &  3.1--10.3 & 0.17$\pm$0.03 & 0.030$\pm$0.005 & -8.310e-06 & 1.2--4.0\,keV\\

continuum - line free regions & 3.10--3.80 & 0.32$\pm$0.08 &                 & -1.340e-06 & \\
          & 4.50--4.89&&&& \\
          & 6.26--6.53&&&& \\
          & 8.53--9.00&&&& \\
          &11.63--12.00&&&& m\\
          &19.20--20.70&&&& m\\
He-like   & 3.88--4.07 & 0.29$\pm$0.06 & 0.046$\pm$0.008 & -2.903e-06 & \\
          & 4.98--5.16 &&&& \\
          & 6.55--6.83+ 5.62--5.73? &&&& \\
          & 9.04--9.40 &&&& \\
          &13.29--13.76&&&& m\\
H-like    & 6.11--6.25 & 0.41$\pm$0.08 & 0.049$\pm$0.019 & -3.411e-06 &\\
          & 8.33--8.50&&&& \\
          &12.03--12.20&&&& m\\
          &18.74--19.13&&&& m\\
Fe-complex 1 &10.31--11.67& 0.35$\pm$0.10 & 0.048$\pm$0.019 & -5.245e-06 & m\\
Fe-complex 2 &14.86--15.52& 0.65$\pm$0.09 & 0.080$\pm$0.028 & -3.457e-06 & m\\
          &16.59--17.22& &&& m\\
\sfir\ line  & 4.98--5.16 & 0.71$\pm$0.16 &                & 5.438e-08 & \\
\hline
\end{tabular}
\end{table*}

\begin{figure*}[!htb]
\begin{tabular}{cc}

\includegraphics[width=3.in,viewport=0 0 571 548,clip=true]{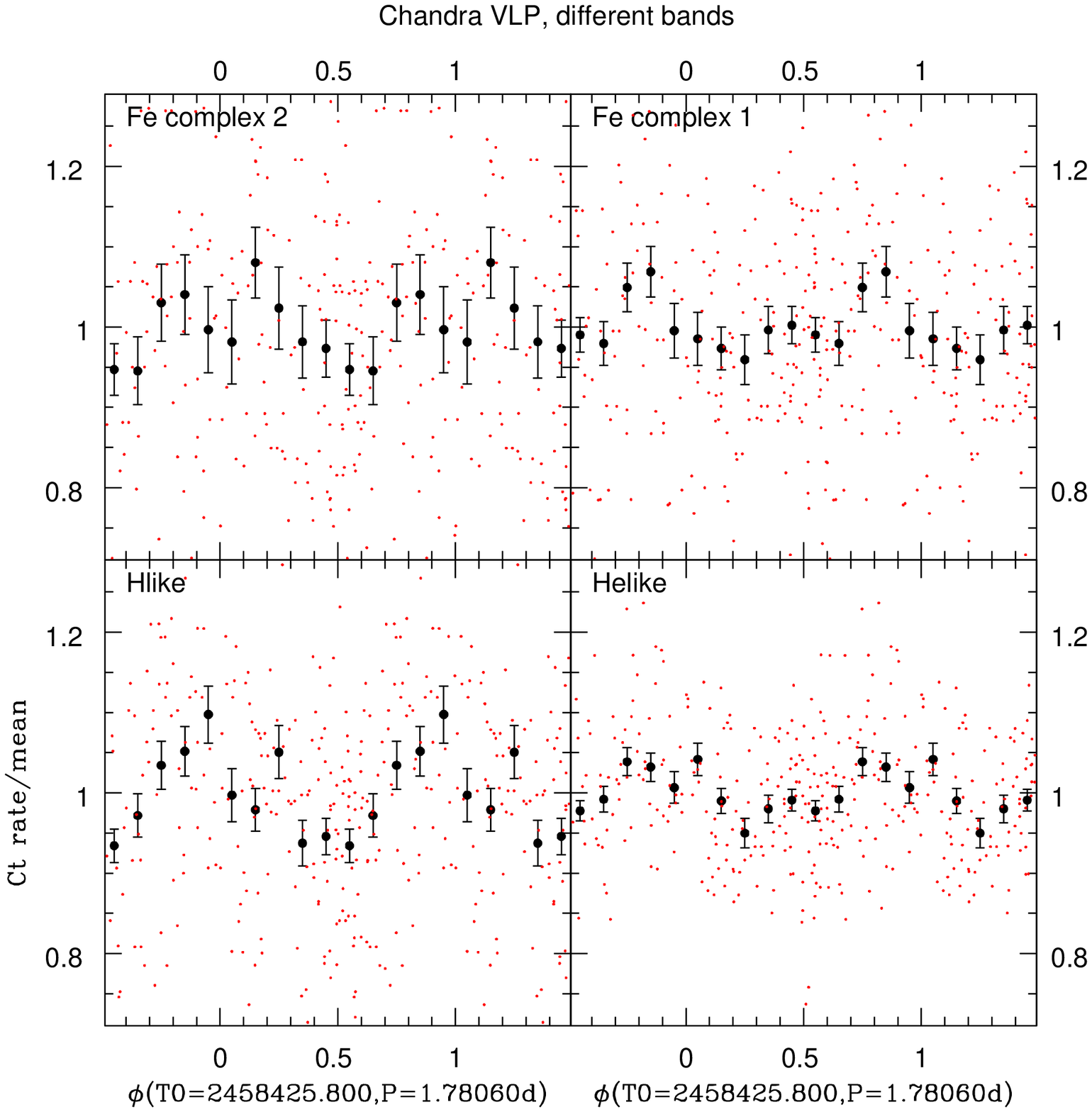} &
 \includegraphics[width=3.in,viewport=0 0 571 548,clip=true]{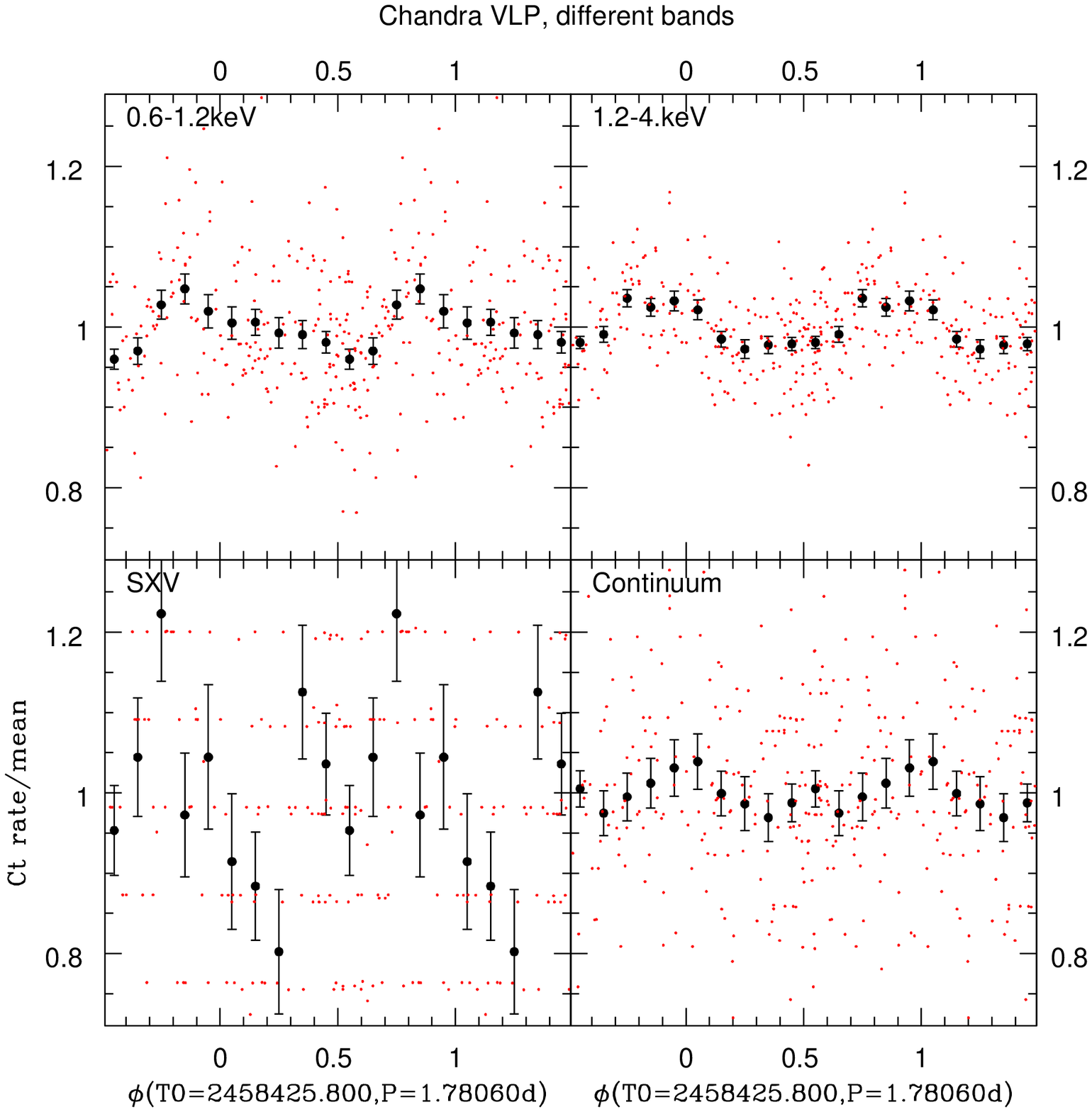} \\
 \end{tabular}
\caption{Binned light curve of  \zp, for several energy bands, folded using the best optical ephemeris and scaled by the average count rate in each band to help visual comparison.  Actual data from the light curves are shown as small red dots.  Two cycles are shown;  data have been replicated in order to show continuity with phase.  The plot for \sfir\  shows the data points in distinct horizontal lines, the result of the very low count rate.}
\label{fig:opt7bc}
\end{figure*}

\section{Spectral region analysis}\label{sec:spectral}
 In addition to the full-band of  \Chandra\ wavelengths, we extracted from each \obs\ narrow-band light curves that include specific spectral regions.  These light curves  contain (1) the sum of  H-like lines of  \siH\, \mgH, \neH, and  \oH, (2) the sum of He-like lines of \arfir, \sfir, \sifir, \mgfir, and \nefir, (3) continuum (selected line-free regions), (4) \Chandra\ soft wavelength bin (10.3--20.7\,\AA), (5) \Chandra\ hybrid-hard  wavelength bin (3.3--10.3\,\AA),  and (6) two Fe complexes, each containing a mix of Fe ionization states from Fe {\textsc {xvii}} to very weak Fe\,\textsc{xx} -- Fe \,\textsc{xxiii}.   The definition of these bands is shown in Table \ref{tab:bands}.   Plots of the light curves of these specific wavelength regions are shown in Fig.\,\ref{fig:opt7bc} .
 These light curves were corrected for the response degradation described  in Sect. \ref{sec:obs} (see slopes in Table\,\ref{tab:bands}).
The light curve analysis of these other energy bands reveals a significant peak near 1.78\,d.  The larger noise makes detection much more difficult in these narrower bands, compared to the full-band pass.
The hardness ratio using the algorithm HR=((hard-medium)/(hard+medium)) with error propagation produced large errors in this noisy data.  There is no evident variability in the hardness ratio, considering the errors, so no plot is shown.

Folding these light curves on the 1.7806\,d period, the peak at $\sim\phi=0.9$ appears significantly detected for Fe complex 1 and H-like lines, and less so for He-like lines, \Chandra\ soft, and \Chandra\ hybrid-hard bands. These binned light curves demonstrate the presence of different behaviors (variation in amplitude and phase evolution) between the energy bands. The binned curves display a coherent trend with phase, although the low count rate for the second  Fe complex and the \sfir\ lines  produce unreliable results. Peak-to-valley amplitudes range from 6 to 16\%. In particular, the H-like lines curve varies, from peak-to-valley, by 16\% while the He-like lines curve only changes by 9\%; the \Chandra\ soft band curve varies by 9\% while that of the hybrid hard band changes by 6\%.  It's not surprising that the different emission line bands can be folded coherently with the 1.78\,d period, as most of the flux in the broad-band light curve comes from the emission lines.  Regarding curve shapes, the \Chandra\ soft band appears somewhat asymmetric, with a steep increase followed by a long, shallower decrease; the curves appear much more symmetric for the hybrid-hard band.

\section{Time sliced spectra} \label{sec:time_sliced_spectra}

To examine variability in fully calibrated spectra, each  \obs\ in Table \ref{tab:obs} was split into multiple pieces using a time filter.  The spectral data were then extracted from each piece as though it were a separate observation.    The time intervals are generally 9\,ks in length, chosen because that time interval fit neatly into most of the exposure times of the \obss, leaving the fewest shorter time slices at the end of each \obs.  Also, 9\,ks provided enough counts in most cases for trend analysis.  Ninety-one time intervals resulted from the time-slicing.  The same products that are available in TGCat for a full  \obs\ were created for each individual time slice.   In particular, Ancillary Response Files (ARFs) and Redistribution Matrix Files (RMFs) were produced and applied for each time interval spectrum.  For all analyses described here, the background was not subtracted or otherwise considered in the flux calculations.  The background, normally measured in two regions parallel to and offset from the spectral arms, is extremely low above 2\,\AA\ relative to the source counts in HETG data ($\le$ 1 ct per extraction cell per Ms) and can be neglected.

\subsection{Dynamic Spectra}

\begin{figure*}[!htb]
  \centering
  \includegraphics[width=0.8\textwidth,angle=270]{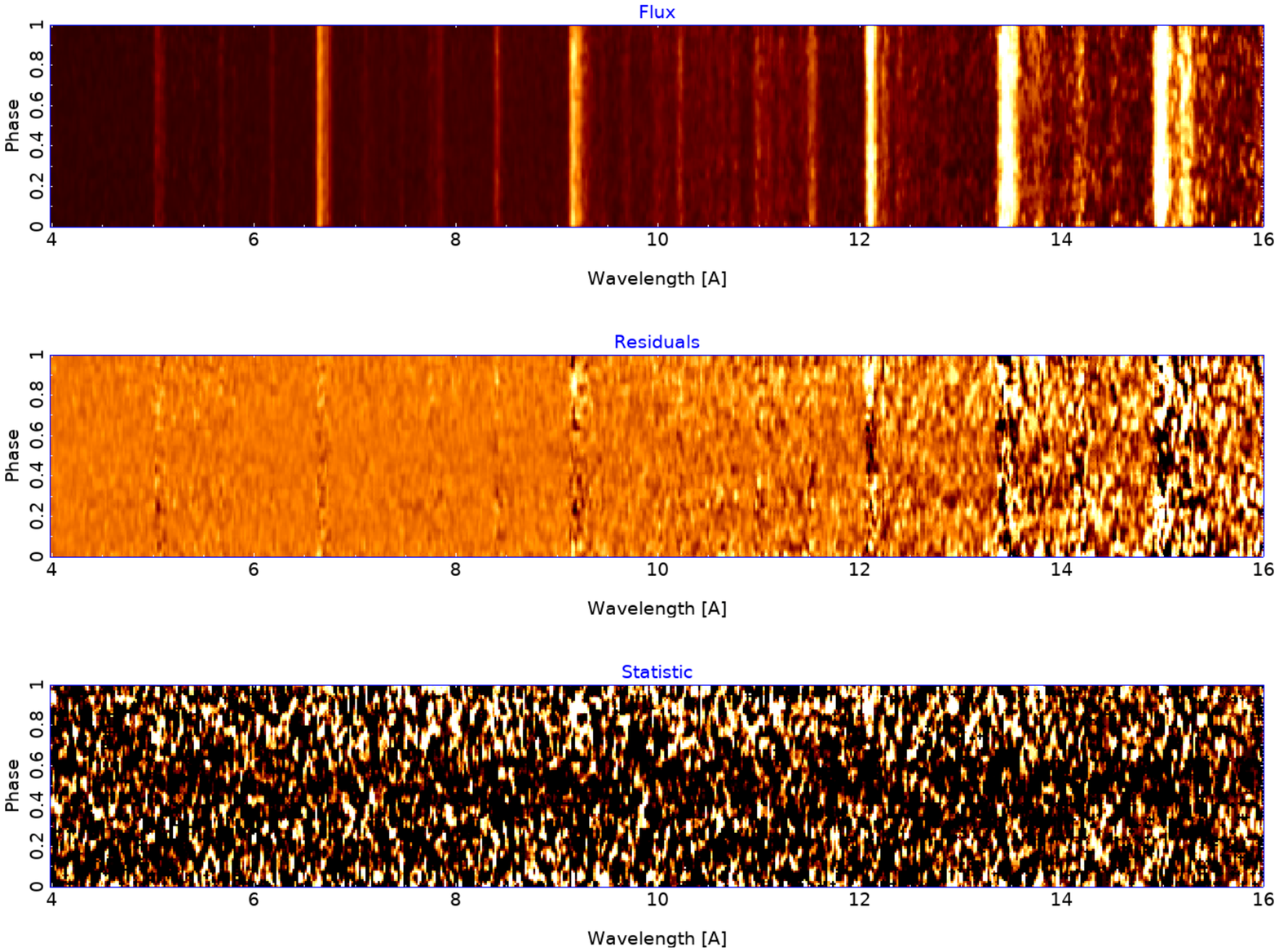}
  \caption{
We show the $\zeta\,$Pup spectrum accumulated into phase bins,
using the period 1.7806\,d, as intensity images in
several forms.  The top panel shows the flux in a linear intensity
scale, with white being high flux and black low flux.  Line identifications are as in Fig.\, \ref{fig:spectrum}. The emission
lines stand out as bright vertical bars.  Using the spectrum summed
over phase as a model, we show the residuals of each phase to that
model in the center panel.  Here we can easily see the flux decrease
in the strong lines near phases $0.5$--$0.7$ as a black, or darker
vertical region (e.g, in \neH\ 12.132\,\AA).  The increase in
variance toward longer wavelengths is due to the decreasing signal,
due to falling effective area with increasing wavelength in this
region.   The bottom panel shows an image of the
residuals divided by the uncertainty, as determined from the counts in
each bin. The scales in the images (top to bottom) are $0$--$5.5\times10^{-3}\,\mathrm{photons\,cm^{-2}\,s^-1}$ (flux),
$-1.0\times10^{-3}$ to $1.0\times10^{-3}\,\mathrm{photons\,cm^{-2}\,s^-1}$ (flux residuals),   and $-1.5$--$0.84$ ($\Delta\chi^2$). One can clearly see the flux decrease in mid-phases
across the entire spectrum.   Using an approximate model for the
spectrum, we have verified that these trends and statistical
fluctuations are qualitatively reproduced with simulated data having
the same counting statistics as the observation, if we impose a 6\%
peak-to-peak amplitude modulation of the flux with phase.}

  \label{fig:dave}
\end{figure*}

To aid in visualization of the time variability of the emission line fluxes, Fig.\,\ref{fig:dave} shows a  period-folded, "trailed" spectrum where the vertical axis represents phase of a 1.7806\,d period, using the ephemeris in Table \ref{tab:ephem}.   All 9\,ks, time-sliced and calibrated spectra were used in constructing the dynamic image.  These images and residuals allow one to view the variations across the spectrum, including emission lines and continuum.  Details are in the caption of Fig.\,\ref{fig:dave}.

\subsection{Moment Analysis} \label{sec:moments}
The \Chandra\ data have modest resolution and significant noise.  Line moments of order zero to two were thus estimated for strong isolated lines (\siH\ 6.182\,\AA, \mgH\ 8.421\,\AA, \neH\ 12.132\,\AA, Fe complex near 11.5\,\AA, and Fe complex near 15.01\,\AA)\footnote{Zeroth-order moment represents flux, first-order moment represents line position, second-order moment represents the square of the line width.}. This was done in both real and simulated 9\,ks slices.

We evaluated the significance of the results of moments by creating a set of ``simulated"
data using the same time filter and calibration files as the real data, with a constant model corresponding to the mean of the full data set and Poisson noise applied randomly.  We then examined these data with the same techniques
as the real data.  Because they will have the same gross statistical properties and window function, analysis of the simulated data in parallel with the actual data gives us another tool to evaluate the reality of any variability we find.
$\chi^2$ tests against constancy reveal the moments to be in general significantly (i.e. $SL<1$\%) variable in real spectra, but not in the simulated dataset, except for line width. Therefore, stellar variability seems to be present at least in flux and centroid values. When phased with the 1.7806\,d period, no coherent flux variations are obvious, but this is probably due to the large noise when considering a single line.

To further assess the variations, we derived the cumulative distribution functions of the moments, both for real and simulated data, and tested their difference using a Kolmogorov-Smirnov test. No significant difference between simulated and real data was found with this test. We also determined the Pearson correlation coefficients between moments of the same lines, again for both simulated and real data. These coefficients never reach values beyond 50\% and there is no systematic difference between simulated and real data (e.g. systematically larger coefficients for real data), hence we rule out the presence of significant correlations between the different moments of a line (e.g. flux-centroid, flux-width, and centroid-width correlations).

Finally, the same period search algorithms were applied to the moments of order zero through 2, for both real and simulated data. No significant difference was found, i.e. peaks with similar amplitudes are found in both datasets. In this case, moment analyses appear too insensitive to allow detection of periodicities in such noisy data.

\section{Discussion} \label{sec:discussion}

While the core of this paper is the observational timing analysis presented above, this section will describe possible links between the measured variability and its physical causes.  Long-term monitoring studies of this star at high cadence and in additional wavebands, as well as detailed modeling, will be needed to make conclusive determinations of the causal mechanisms for the observed variability.

\subsection{The 1.78 Day Period}  \label{subsubsec:Obs1.78}

In the full spectral range, a period of 1.7820\,d is clearly identified in the X-ray data and is almost certainly closely connected to the previously-measured optical modulation with essentially the same period.   The peak-to-valley amplitude in our observations for the P = 1.7820\,d signal is about 6\%, which is remarkably large compared to the optical variations of only 1\%.
The 1.7820\,d period we found in the \Chandra\ full-band data is not peculiar to some small region of the X-ray spectrum, but is relatively general through the whole range of wavelengths (see Fig.\,\ref{fig:dave}).  It is present in the flux from the H-like lines, He-like lines and a collection of Fe lines.  It even appears to be present in the portions of relatively line-free continuum.  The light curve characteristics however appear slightly different, both in shape and amplitude, for different bands or lines.  A closer investigation of the emission lines (though strongly limited by the small number of photon counts in each line) did not reveal clear periodic changes in line properties other than flux.  Moment analysis of a number of emission lines individually does not indicate any correlation between line width and centroid velocity.

Considering only the maxima in the optical ($\phi$=0.45) and X-ray ($\phi$=0.9) light curves, the lag of $\phi\sim= 0.45$ of a cycle in phase might provide an important clue in untangling the connection between the two bands.   Assuming the maximum X-ray flux lags the maximum optical peak is a possible interpretation of the relation between the optical and X-ray light curves due to the amplitudes of the maxima.  However, it is  also possible  that the X-ray maximum is actually physically related to the secondary optical maximum because the two phenomena occur at almost equivalent phases (secondary optical: $\phi=0.8$; primary X-ray: $\phi=0.9$) and the lag time would be $\sim \phi=0.1$.  Also, the secondary maximum in the \Chandra\ light curve may be related to the \BRITE\ maximum (primary optical: $\phi$=0.45; secondary X-ray: $\phi$=0.55), again because they are at nearly the same phase.  Together, the lag time for this set of features would each be $\sim \phi$=0.1.

\subsection{The Period of 5\,d or 6\,d}  \label{subObs5-6}

In addition to the main 1.7820\,d X-ray period, there is an indication of some periodicity close to 5\,d or to 6\,d.  This periodicity is more difficult to pin down for two reasons: (1) the breadth and complexity of this peak in the periodogram (see Fig.\,\ref{fig:yael}, lower left panel), and (2) the lack of any comparable periodicity in the contemporaneous \BRITE\ data.  Though simply choosing the highest peak in this region of the periodogram would favor a period near 6 days, the nearly equally prominent peak near 5 days (see Fig.\,\ref{fig:joy_506}) is particularly interesting because of its proximity to two previously measured periodicities for this star.   \citet{moffat81} claimed a 5.1\,d periodicity in the near-central reversal in the H$\alpha$ emission line, and explained it by excess magnetically-confined plasma.  \citet{howarth95} found a 5.2 $\pm$ 0.7\,d period in two solid uninterrupted weeks of \IUE\ UV spectra, described as being of unknown origin.
The light curve of \Chandra\ data folded on a 5.056\,d period (Fig.\,\ref{fig:joy_506}) suggests that this period is possible, although not as convincingly as the 1.78\,d period.

\subsection{Stochastic Variability}   \label{sec:stochastic}

Finally, a residual stochastic component to the X-ray variability cannot be ruled out. After removing the two periodicities described above from the signal, the scatter in the residual broad-band X-ray light curve is $\sim$6\%, compared to the estimated Poisson noise of 4\%. Since there should not be any significant additional instrumental sources of noise, this leaves a net rms scatter of $\sqrt{(6^2 - 4^2)} =\sim$ 4.5\%, which, if real, would be intrinsic to the wind. However, as shown above from the $F_\mathrm{var}$ determination, the presence of stochastic variability is only a 1.2$\sigma$ detection; hence the need for confirmation.  If confirmed, this might be understood in terms of stochastic shocks throughout the wind leading to random fluctuations at this level in the X-ray flux.   The X-ray modulations are 0.03 for the initial dataset, 0.02 after cleaning for one period, and 0.01 after cleaning by both periods, i.e. the 1.78\,d and 6\,d periods may have similar strengths as the remaining variability. A similar ratio of periodic vs stochastic variability was also found in the optical \citep{ramia18}.
Nevertheless, because of the Poisson noise,
additional data are clearly needed to clarify the presence of
these stochastic variability.

Assuming for the moment that this stochastic component
exists and will be confirmed in the future, one is thus drawn
to the idea that we are seeing similar effects in X-rays as in
optical data regarding CIRs and wind clumps. If the X-ray
emission showed no intrinsic stochastic emission, it would indicate
a scenario of Poisson saturation due to a myriad of clumps.   \citet{naze13} found that the \zp\ wind should contain at least 10$^5$ X-ray-emitting shocks, leaving a relative Poisson fluctuation of $\sim 1\sqrt{10^5} =$ 0.3\%, well below detection limits of current X-ray telescopes. However, the modeling for this estimate neglects the more likely scenario of a turbulent powerlaw with progressively fewer clumps or shocks of large scale \citep{moffat94}. Such rare large clumps/shocks also emit and scatter more photons than their smaller cousins, which may lead to detected stochastic fluctuations, as now seen in \zp\ in the optical with \BRITE, and possibly in X-rays in the observations described here.

\subsection{Periodic Variability: Physical Mechanisms}
\label{sources_of_variability}
When trying to determine the physical origin of some stable monoperiodicity in the signal received from a star, rotation is of course the most obvious culprit. In addition, the fact that we have now clearly seen the same period in two wavelength bands has other ramifications.  If rotation is accepted as the principal motor driving a periodic behavior of the star there must be some way of connecting surface inhomogeneties in the optical photosphere with some sort of modulation of the X-ray emitting and absorbing regions of the wind.   This makes it very tempting to invoke CIRs, with the optical (continuum) light curve arising from the bright spot on the stellar surface at the base of the CIR and the X-ray light curve arising from time-varying visibility of shocks in the CIR due to its rotational modulation out in the wind. Qualitatively, this idea fits with intermediate phase lags up to 0.1 in  phase seen in three different successive optical recombination lines compared to the continuum light coming from the photosphere \citep{ramia18}.  There are two possible mechanisms that could cause a CIR imprint on the X-ray signal:   (1) as a CIR sweeps through the un-occulted portions of the wind, rotation could modulate the X-ray signal by revealing zones of additional emission, or (2) obscuration of the shocks could be caused by increasingly more or less wind material between the X-ray sources and the observer.
If CIRs are indeed causing the modulation of the X-ray signals, there is an interesting interplay between the inclination angle of the star and the physical structure of CIRs.  Obviously, if the star is seen exactly pole-on, there would be no rotational modulation of the signal.  If the inclination angle of the star could be independently constrained, that knowledge would reveal information about the extent of CIRs in both radius and latitude.  At small inclination angles, only X-ray-emitting material very near the star would be noticeably occulted each rotation.

In addition to the X-ray modulation described here, DACs, thought to be an observational manifestation of CIRs,  have been observed in the past for this star in the UV.  This independent line of evidence indicates that the inclination of \zp\ must be far enough from pole-on to allow some rotational modulation of the signal.   If a low inclination means our line of sight to this star may just clip the high-latitude portions of the CIRs, this may explain why DACs have been more variable in this star than in some other similar stars.

It should be noted that while it is difficult to explain any regular periodicity with periods longer than the rotational period, it is relatively easy to explain any shorter period harmonically related to the rotation period of the star.  Physically, this would be manifested as having multiple (more or less persistent) structures at different longitudes around the star. If, for instance, there are two CIR footpoints (``hot spots"), the rotation period P$_\mathrm{rot}$ could be 1.78\,d or 2 x 1.78\,d = 3.56\,d.  This configuration (two structures per 2$\pi$) has been discussed as being plausible  by \citet{kaper99}, \citet{dejong01}, and \citet{Massa19}, and is consistent with the period we determined of 1.7820\,d.  However if the period is 3.56\,d, the two structures must be sufficiently symmetrical that the true rotational fundamental of 3.65\,d doesn't show up strongly in our data or in optical data.  Further knowledge of this star's CIRs is needed to gauge the likelihood of this scenario of very symmetric CIRs separated by exactly 180 degrees.
Two rotation possibilities related to the 1.78\,d signal can be summarized as follows:
\begin{itemize}
\item  If P$_\mathrm{rot}$ = 1.78\,d, the star would rotate very close to break-up, which poses questions regarding the wind driving, wind symmetry and the lack of a disk.  A star so near break-up would be somewhat oblate.  One or more hot spots could be present.
\item If P$_\mathrm{rot }= 2 \times 1.78$\,d, the question is why this period does not show up in the periodogram above noise level. A significant degree of symmetry would be needed between the two hemispheres.  This scenario would imply rotation velocity safely below critical and perhaps a more equator-on view.  While this rotation period would still comply with the limits posed by $v \sin i$ and distance, the fit is not as comfortable as for the 1.78\,d period.
\end{itemize}

It might be difficult to discriminate between these two cases because both of these possible rotation periods fit (if just barely for the longer one) in the range of allowable rotation periods in a recent detailed analysis of the fundamental properties of this star.  \citet{howarth19} argue that the ``revised" Hipparcos distance of d = 332 $\pm$ 11 pc is reliable.  Using this as their basis, their analysis went on to exclude any period above 3.7\,d with 95\% confidence. The argument was based on the robust spectroscopic measurements of $v\_eq\times \sin(i) = 213 \pm 7$ km/s, and the obvious maximum of $sin(i)=1$ (equator-on view).

\citet{howarth19} discuss the consequences of the surprisingly small Hipparcos distance on \zp\  physical properties,  namely that it would have a relatively low luminosity and thus mass, atypical for its spectral type.  If \zp\ in fact has more normal properties for its spectral type, its true distance is larger and a rotation period of 2$\times$1.78\,d is plausible within the limit posed by the measured $v \sin i$.

If we relax some of these assumed constraints, it is within the realm of possibility that the 5\,d or 6\,d period apparent in our data is the true rotational period of this star.  We however consider this unlikely for the following reasons.  First, to have such a long rotational period, the distance to the star must be very much greater than that found by \citet{howarth19} as described above.  Second, it is not clear why the 1.78\,d signal would be so strong when it would just be a harmonic of the fundamental (rotation) period.  To hypothesize a rotation period in the 5\,d or 6\, d range, the 1.78\,d period would need to be the n=3 harmonic of the fundamental, indicating that the true rotation period would be 5.35\,d.   The possibility of the 1.78\,d period being the n=3 harmonic is discussed in Sect.\,\ref{sec:period}.  If the 1.78\,d period were the 4th harmonic of the rotational period, the rotational period would have to be 7.12\,d, outside the range of periods found on the periodogram, and this is of course strongly excluded under the recent distance determination.

To combine these effects into a specific physical example, let us say that the rotational period of the star is indeed three times the 1.78\,d period (5.35\,d).  This would imply a distance of 480\, pc or more, which does not look likely in light of the recent work of  \citet{howarth19} described above.
We therefore conclude that the ``rotation period" (as defined above) is most likely either $1 \times 1.78$\,d  or $2 \times 1.78$\,d.

Finally, there is a note of caution which should be applied when comparing periods (and harmonics) measured for an individual star using different methods and at different epochs.  There is no \textit{a priori} reason why hot stars should be solid body rotators. \citet{howarth19} proposed anti-solar differential rotation. Structures at different latitudes could be going around at slightly different angular speeds.  Alternatively, a structure causing variability in some specific waveband could migrate in latitude over time, causing a change in the measured rotation period.  In our case, $P_\mathrm{rot}$ means the rotation period at the latitude where the structures are located that give rise to the observed variability at the epoch of observation.

When evaluating possible physical explanations for these patterns of variability, it is useful to review the evolutionary history of \zp.  \zp\ is currently believed to be a single, massive runaway star.  When it was in a binary system with a more massive primary star,  the Roche Lobe Overflow (RLOF) process would have spun up \zp\  in the time before the primary's SN explosion.  After that explosion, the secondary star (now \zp) was ejected from the system with high spin-rate in the opposite direction to that of the remnant compact primary.  It therefore would not be surprising to find that \zp\ shows rapid rotation.    It was noted above that accepting a relatively close distance for \zp\ would indicate that it is under-luminous for a star of its spectral type, but it is possible to explain this discrepancy by appealing to its individual evolutionary history in a mass-exchange binary.

The putative evolutionary history of \zp\ could conceivably make a different contribution to the periodic variability of this system.  There is some finite possibility that \zp\ remained bound with at least a part of the debris from the exploding star or material participating in the RLOF.  The 5\,d or 6\,d periodicity could be caused by some sort of low-mass companion object (of whatever origin) orbiting \zp\ at a distance of $\sim$3 stellar radii.  Rotation is the obvious prime mover for any clock-like periodicity for this star, but orbital motion provides many other options.   Orbital motion could explain any stable periods longer than the rotational period.  Such orbital motion periods would be expected to have no specific relation to that rotational period.  While this scenario is highly speculative, we mention it for the sake of completeness.

An additional source of periodicity could be Non-Radial Pulsations  (NRPs).  As discussed in \citet{howarth19},   \citet{howarth14} had applied the theoretical pulsation models of \citet{siao11} to this star.   Acceptance of the new nearer distance, with its attendant under-luminosity, and the individual evolutionary history of \zp\, make it difficult to apply the \citet{siao11} models which are for standard single-star evolutionary tracks.  Without detailed modeling based on individual properties of \zp\ giving specific modes, periods, and amplitudes, it is not possible to evaluate what contributions these pulsations make to the observed periodic variabilities.

\subsection {Discussion of phase difference with respect to CIR parameters}
\label{lag}

Setting aside the determination of the causes of the specific values of the periodicities in the observation, the other extremely interesting aspect of the data is the phase difference between the flux maxima in the X-ray and optical folded light curves.  This paper will not attempt detailed modeling, but from a theory perspective a most important clue to the nature of the variability is the fact that most of the roughly 6\% (peak-to-peak) X-ray variation is coherent on a period of 1.7806\,d  over a time of up to a year.  If we assume that the maximum of the \BRITE\ light curve at $\phi$=0.45 is related to the \Chandra\ light curve maximum at $\phi$=0.9, the time lag is about $6.1 \times 10^4$\,s  relative to the \BRITE\ maximum. If the 1.78\,d period is interpreted as a rotation period (as in \citet{ramia18}), this corresponds to a phase lag of some 45\% of a cycle between the optical maximum and the X-ray peak. A potential interpretation for this phase lag is the curved shape of a CIR in the wind \citep{cranmer96}, caused by the effects of rotation on wind streams with different rates of radial acceleration.

Analyzing the coriolis influence on wind acceleration in the corotating frame shows that the radius of a CIR shock, in stellar units, caused by a bright photospheric spot \citep{cranmer96}, is characterized by the ratio of the terminal speed to the rotation speed, times the angular scale (in radians) of the spot on the stellar surface.   The phase lag of the X-ray hot spot where streamlines converge is of the same order as the size of the spot.  The spot brightness contrast must be fairly significant in order to produce a strong shock (by overloading the local mass flux, it stalls and gets rammed from below).  Since the optical brightness variations are only at the 1\% level, it means the spot must be relatively small, no more than a few percent of the stellar surface and covering a phase interval no larger than 0.1 of the rotation period.  For \zp\ the ratio of terminal speed to rotation speed is about 10, so the shock forms at much less than 10 stellar radii, possibly even in the range 1-3 where we also expect the bulk of the X-rays to form.   Because small spots and low X-ray radii are consistent with a lag in the X-ray emission of no more than about  0.1 of the rotational period, we would have difficulty forming a consistent picture if we thought the X-ray peak was associated with the \BRITE\ peak about $\phi= 0.45$  earlier in the rotational period.  Thus, either the X-ray peak is associated with the weaker \BRITE\ peak near phase 0.8, or the \BRITE\ signal is created by a dark spot offset by $\phi=0.5$ from our expectation.  Although dark spots are equally capable of producing X-ray shocks, because the latter only require a change in terminal speed and dark spots can underload the wind and generate fast streams,  \citet{cranmer96} found that dark spots do not generate DACs.  Hence, the most self-consistent interpretation is that the X-ray peak is associated with the second, albeit weaker, \BRITE\ peak.  What would be necessary to verify this interpretation is simultaneous X-ray, optical and UV line observations, to test the rotational phase relationships of DACs and CIRs relative to X-ray generation.

Thus, it seems more likely that the two optical light curve maxima represent two different surface hot spots on the star.  The \Chandra\ light curve also has two potential maxima. The similarity between the phase lags for each of these X-ray/optical pairs suggests that, rather than a phase difference of $\phi$=0.45 for the primary maxima only, there are two optical features, each with an associated X-ray peak and a lag time of $\sim \phi=0.1$.  This value is similar to lag times for other stars comparing X-ray and multiwavelength light curves \citep{Massa19}.  If this is in fact the case, the situation is that the largest optical maximum is associated with a rather small secondary maximum in X-ray, while the secondary optical maximum is associated with the largest feature in the X-ray light curve.
We can perhaps understand such a configuration if we consider that the X-ray flux, as mentioned above, depends on the viewing angle of the curved CIR in the wind and  on occultation.  Thus, the strength of the X-ray signal may not be clearly correlated to the structure of the optical emission from the hot spot itself.
A DAC period of about 20\,h could be explained if the light curves are interpreted as displaying evidence for two hot spots.  DAC periods might fit into this scenario with  periods of about 0.8\,d which is about half of the rotation period.

\newpage
\section{Conclusions}\label{sec:conclusions}

Using the large dataset of Chandra HETG observations of \zp\, we have identified a 1.7820\,d period in the X-ray data that is within the errors of the 1.7806\,d period identified in optical observations.  The maximum of the X-ray light curve is out-of-phase with the optical maximum by $\sim\phi=0.45$ in phase.
However, if the secondary maxima in the optical and X-ray are considered, the phase lags for these two hot spots/CIRs complexes are about $\phi$=0.1 each.  In addition, a secondary period of 5\,d or 6\,d, although marginally detected, may be consistent with some previous UV and optical periods.  The data are not inconsistent with, but cannot definitively confirm, the presence of intrinsic stochastic variability.  We have explored in detail the difficulties of  accepting as the rotation period  either 1.78\,d, 2x1.78\,d, 5\,d ,or 6\,d, but conclude that the rotation period is most likely 1.78\,d.  Finally an attempt was made to explain the time lag in X-ray and optical light curve maxima.  A preliminary calculation shows that, assuming the maximum X-ray emission is formed in the CIR curve, the lag time determined from the observations of $\phi$=0.45  implies a formation position too many stellar radii from the stellar surface to be plausible with current theory.  Rather, the possibility of the detection of two hot spots on the star with X-ray emission in the curved CIRs is considered more likely.  The new observational phenomena presented in this paper will need significant modeling efforts.

In summary, \zp\ is now a source with a number of clearly established periodicities, some of which display interesting links across multiple wavebands.  Though the physical origin of these variations is still somewhat unclear, the rich data set being developed for this star indicate the usefulness of variability analysis as a probe of connections between the photosphere and wind.  Future long term, intensive, multiwavelength photometric and spectroscopic monitoring of this important astrophysical source is certainly warranted.

\section{Acknowledgements} JSN acknowledges a grant from the \Chandra\ X-ray Center to support a research visitor for this project.  JSN also acknowledges support of  \Chandra\ contract  NAS8-03060.  Chandra General Observer Program, Cycle 19 supported WW by GO8-19011A, JSN by GO8-19011B, NAM by GO8-19011D, NDR by GO8-19011E, and RI by GO8-19011F.  Y.N. acknowledges support from the Fonds National de la Recherche Scientifique (Belgium), the European Space Agency (ESA) and the Belgian Federal Science Policy Office (BELSPO) in the framework of the PRODEX Programme (contract HERMES).  AM is grateful for financial aid from NSERC (Canada).  TR acknowledges support from the NASA APRA program (NNH16ZDA001N-APRA).  NAM also acknowledges support from the UWEC Office of Research and Sponsored Programs through the sabbatical and URCA programs, and from a Chandra Research Visitor award.  NAM would like to thank Jacob Richardson for useful initial discussions of the data.
All authors wish to acknowledge the assistance of the Northrop Grumman \Chandra\ Flight Operations Team for innovative work in acquiring these observations and dedication to the project.  This research has made use of data obtained from the \Chandra\ Data Archive, and software provided by the \Chandra\ X-ray Center (CXC) in the application packages CIAO, ChIPS, and Sherpa.  This research has made use of ISIS functions (ISISscripts) provided by
ECAP/Remeis observatory and MIT (http://www.sternwarte.uni-erlangen.de/isis/).

\software{TGCat \citep{Huenemoerder11}},{CIAO \citep{Frusc06}}
\facilities{CXO,BRITE}
\bibliographystyle{apj}
\bibliography{zetapup_variability}

\end{document}